\documentclass[prd,aps,onecolumn,superscriptaddress,nofootinbib,showpacs,preprintnumbers,longbibliography]{revtex4-1}

\usepackage{amsmath}
\usepackage{amssymb}
\usepackage{bm}
\usepackage{multirow}
\usepackage{float}
\usepackage{bbold}

\usepackage{graphicx}
\usepackage[american]{babel}
\usepackage{enumerate}
\usepackage{xcolor,colortbl}
\usepackage{mathtools}
\usepackage{booktabs}
\usepackage{subcaption}
\usepackage{soul}
\setstcolor{red}
\usepackage{makebox}
\newcommand{\U}[1]{\mathrm{U}(1)_{\mathrm{#1}}}			
\newcommand{\SU}[2]{\mathrm{SU}(#1)_{\mathrm{#2}}}		
\newcommand{\SO}[2]{\mathrm{SO}(#1)_{\mathrm{#2}}}		
\newcommand{\E}[1]{\mathrm{E}_{#1}}	
\newcommand{\g}[2]{g_{_\mathrm{#1}}^{#2}}

\renewcommand{\(}{\left(}
\renewcommand{\)}{\right)}
\renewcommand{\[}{\left[}
\renewcommand{\]}{\right]}
\newcommand{\del}{\partial}
\newcommand{\abs}[1]{\left| #1 \right| }
\newcommand{\mean}[1]{\left \langle #1 \right \rangle }
\newcommand{\vev}[0]{VEV}
\newcommand{\vevs}[0]{VEVs}

\newcommand{\ro}[1]{\textrm{#1}}
\usepackage{pifont}
\newcommand{\xmark}{\ding{55}}%

\graphicspath{{figures/}}

\allowdisplaybreaks

\begin{document}

\title{What can a heavy $\U{B-L}$ $Z^\prime$ boson do to the muon $\left(g-2\right)_\mu$ anomaly and to a new Higgs boson mass?}

\author{Ant\'onio P. Morais$^{a,b}$} 
\email{aapmorais@ua.pt}
\author{Roman Pasechnik$^{b}$} 
\email{Roman.Pasechnik@thep.lu.se}
\author{J.~Pedro Rodrigues$^{a}$} 
\email{joaopedrorodrigues@ua.pt}

\affiliation{
{$^a$\sl 
Departamento de F\'{i}sica da Universidade de Aveiro and CIDMA
	Campus de Santiago, 3810-183 Aveiro, Portugal
}\\
{$^b$\sl 
Department of Astronomy and Theoretical Physics,
Lund University, SE 223-62 Lund, Sweden
}}


\begin{abstract}
\vspace{0.5cm}
The minimal $\U{B-L}$ extension of the Standard Model (B-L-SM) offers an explanation for neutrino mass 
generation via a seesaw mechanism as well as contains two new physics states such as an extra Higgs boson and 
a new $Z'$ gauge boson. The emergence of a second Higgs particle as well as a new $Z^\prime$ gauge boson, both 
linked to the breaking of a local $\U{B-L}$ symmetry, makes the B-L-SM rather constrained by direct searches at 
the Large Hadron Collider (LHC) experiments. We investigate the phenomenological status of the B-L-SM by confronting 
the new physics predictions with the LHC and electroweak precision data. Taking into account the current bounds 
from direct LHC searches, we demonstrate that the prediction for the muon $\left(g-2\right)_\mu$ anomaly in 
the B-L-SM yields at most a contribution of approximately $8.9 \times 10^{-12}$ which represents a tension 
of $3.28$ standard deviations, with the current $1\sigma$ uncertainty, by means of a $Z^\prime$ boson 
if its mass lies in a range of $6.3$ to $6.5~\ro{TeV}$, within the reach of future LHC runs. This means
that the B-L-SM, with heavy yet allowed $Z^\prime$ boson mass range, in practice does not resolve 
the tension between the observed anomaly in the muon $\left(g-2\right)_\mu$ 
and the theoretical prediction in the Standard Model. Such a heavy $Z^\prime$ boson also 
implies that the minimal value for a new Higgs mass is of the order of 400 GeV.
\end{abstract}

\maketitle

\section{Introduction}
\label{sec:Introduction}

It is unquestionable that the Standard Model (SM) is a successful framework accurately describing the phenomenology of Particle Physics up to the largest energy scales probed by collider measurements so far. In fact, contemporary direct searches for new physics or indirect probes via e.g.~flavour anomalies, have been showing an increasingly puzzling consistency with SM predictions. However, it is not less true that the SM also possesses its weaknesses and several open questions are yet to be understood. One of such weaknesses is a missing explanation of tiny neutrino masses confirmed by flavour-oscillation experiments. The minimal way of addressing this problem is by adding heavy Majorana neutrinos in order to realise a seesaw mechanism \citep{Yanagida:1979as,GellMann:1980vs,Mohapatra:1979ia}. However, the mere introduction of an arbitrary number of heavy neutrino generations can raise new questions, in particular, how such a new scale is generated from a more fundamental theory. 

Among the simplest ultraviolet (UV) complete theories that dynamically addresses this question is the minimal gauge-$\U{B-L}$ extension of the SM \cite{Davidson:1978pm,Mohapatra:1980qe,Basso:2010hk,Basso:2011na}, traditionally dubbed as the B-L-SM. As its name suggests, the B-L-SM promotes an accidental conservation of the difference between baryon (B) and lepton (L) numbers in the SM to a fundamental local Abelian symmetry group. Furthermore, such a $\U{B-L}$ symmetry can be embedded into larger groups such as e.g.~$\SO{10}{}$ \cite{Chanowitz:1977ye,Fritzsch:1974nn,Georgi:1978fu,Georgi:1979dq,Georgi:1979ga} or $\E{6}$ \cite{Achiman:1978vg,Gursey:1975ki,Gursey:1981kf}, making the B-L-SM model well motivated by Grand Unified Theories (GUTs). The presence of three generations of right-handed neutrinos also ensures a framework free of anomalies with their mass scale developed once the $\U{B-L}$ is broken by the VEV of a complex SM-singlet scalar field, simultaneously giving mass to the corresponding $Z^\prime$ boson.

The cosmological implications of the B-L-SM are also relevant. First, the presence of an extended neutrino sector implies the existence of a sterile state that can play a role of $\ro{keV}$- to $\ro{TeV}$-scale Dark Matter candidate \cite{Kaneta:2016vkq}. Particularly, it can be stabilized by imposing a $\mathbb{Z}_2$ parity as it was done e.g.~in Refs.~\cite{Okada:2010wd,Okada:2016gsh,Okada:2018ktp}. The $\U{B-L}$ model with sterile neutrino Dark Matter can also explain the observed baryon asymmetry via the leptogenesis mechanism (see Refs.~\cite{Fukugita:1986hr,Pilaftsis:1997jf,Pilaftsis:2003gt,Blanchet:2009bu,Dev:2017xry}, for details). As was mentioned earlier, the B-L-SM features an extended scalar sector with a complex SM-singlet state $\chi$ which, besides enriching the Higgs sector with a new potentially visible state, can cure the well-known metastability of the electroweak (EW) vacuum in the SM \cite{Degrassi:2012ry,Alekhin:2012py,Buttazzo:2013uya}. Indeed, it was shown in Ref.~\cite{Costa:2014qga} that an additional physical scalar with a mass beyond a few hundred $\ro{GeV}$ can stabilize the Higgs vacuum all the way up to the Plank scale. In the framework of B-L-SM, a complete study of the scalar sector was performed in Ref.~\cite{Basso:2010jm} where the vacuum stability conditions valid at any Renormalization Group (RG) scale were derived. Last, but not least, the presence of the complex SM-singlet $\chi$ interacting with a Higgs doublet typically enhances the strength of EW phase transition potentially converting it into a strong first-order one \cite{Barger:2008jx}.

Another open question that finds no solution in the SM is the discrepancy between the measured anomalous magnetic moment of the muon, $a_\mu^\ro{exp} \equiv \tfrac{1}{2} \(g-2\)^\ro{exp}_\mu$, and its theoretical prediction, $a_\mu^\ro{SM} \equiv \tfrac{1}{2} \(g-2\)^\ro{SM}_\mu$, which reads \cite{Tanabashi:2018oca,Zyla,Davier:2017zfy,Davier:2019can}
\begin{equation}
	\label{g-2}
	\Delta a_\mu = a_\mu^\ro{exp} - a_\mu^\ro{SM} = 261(63)(48) \times 10^{-11}
\end{equation}
with numbers in brackets denoting experimental and theoretical errors, respectively. This represents a tension of $3.3$ standard deviations from the combined $1 \sigma$ error and is calling for new physics effects beyond the SM theory. In a recent work \cite{Campanario:2019mjh} it was further claimed that SM higher order perturbative corrections cannot explain $\Delta a_\mu$. A popular explanation for such an anomaly resides in low-scale supersymmetric models \cite{Belyaev:2016oxy,Grifols:1982vx,Ellis:1982by,Kosower:1983yw,Yuan:1984ww,Romao:1984pn,Cho:2011rk,Okada:2013ija,Endo:2013lva,Gogoladze:2014cha,Wang:2015rli} where smuon-neutralino and sneutrino-chargino loops can explain the discrepancy \eqref{g-2}. However, this solution is by no means unique and radiative corrections with new gauge bosons can also 
contribute to the theoretical value of the muon anomaly \cite{Czarnecki:2001pv,Appelquist:2004mn,Kang:2019vng,Lindner:2016bgg}. This is indeed the case of the B-L-SM, or its SUSY version \cite{Khalil:2015wua,Yang:2018guw,Cao:2019evo}, where a new $Z^\prime$ gauge boson can explain $\Delta a_\mu$. 

In a recent work \cite{Deppisch:2019ldi}, the impact of LHC searches for a light $Z^\prime$ boson, i.e.~with mass in the range of $0.2~\ro{GeV}$ to $200~\ro{GeV}$, was thoroughly investigated. 
The current collider bounds are available from the ATLAS  \cite{Aad:2019fac} and CMS \cite{Sirunyan:2018exx} searches for Drell-Yan $Z^\prime$ production decaying into di-leptons, 
i.e.~$pp \to Z^\prime \to ee,\mu \mu$. In the current work, we perform a complementary study where, for heavy (TeV-scale) $Z^\prime$ masses, the combined effect 
of the electroweak precision and Higgs observables and collider constraints on the $pp \to Z^\prime \to ee,\mu \mu$ channel, is investigated. We analyse whether the existing LHC constraints leave any 
room for partially explaining the $(g-2)_\mu$ anomaly and which impact it has on the model parameters and other physical observables such as the $\U{B-L}$ gauge coupling $\g{B-L}{}$, the kinetic mixing 
parameter $\g{YB}{}$ and the extra scalar and $Z^\prime$ boson masses. Furthermore, with the current Muon $(g-2)_\mu$ experiment E989 at Fermilab \cite{Grange:2015fou}, it will be possible to either 
confirm or eliminate, at least partially, the currently observed discrepancy, making our work rather timely.

The article is organized as follows. In Section 2, we give a brief description of the B-L-SM structure focusing on the basic details of scalar and gauge boson mass spectra and mixing. In Section~\ref{sec:parameter-space-studies}, a detailed discussion of the numerical analysis is provided. In particular, we outline the methods and tools used in our numerical scans as well as the most relevant phenomenological constraints leading to a selection of a few representative benchmark points. Besides, the numerical results for correlations of the $Z^\prime$ production cross section times its branching ratio into light leptons versus the model parameters and the muon $(g-2)_\mu$ are presented. Finally, Section~\ref{sec:Conclusions} provides a short summary of our main results.

\section{Model description}

In this section, we highlight the essential features of the minimal $\U{B-L}$ extension of the SM relevant for our analysis. Essentially, the minimal B-L-SM is a Beyond the Standard Model (BSM) framework containing three new ingredients: 1) a new gauge interaction, 2) three generations of right handed neutrinos, and 3) a complex scalar SM-singlet. The first one is well motivated in various GUT scenarios \cite{Chanowitz:1977ye,Fritzsch:1974nn,Georgi:1978fu,Georgi:1979dq,Georgi:1979ga,Achiman:1978vg,Gursey:1975ki,Gursey:1981kf}. However, if a family-universal symmetry such as $\U{B-L}$ were introduced without changing the SM fermion content, chiral anomalies involving the $\U{B-L}$ external legs would be generated. A new sector of additional three $B-L$ charged Majorana neutrinos is essential for anomaly cancellation. Also, the SM-like Higgs doublet, $H$, does not carry neither baryon nor lepton number, therefore does not participate in the breaking of $\U{B-L}$. It is then necessary to introduce a new scalar singlet, $\chi$, solely charged under $\U{B-L}$, whose VEV breaks the $B-L$ symmetry at a scale $\mean{\chi} > \mean{H}$. It is also this breaking scale that generates masses for heavy neutrinos. The particle content and charges of the minimal $\U{B-L}$ extension of the SM are summarized in Tab.~\ref{tab:charges}.
\begin{table}[htb!]
	\begin{center}
		\begin{tabular}{ccccc}
			& $\SU{3}{C}$ & $\SU{2}{L}$ & $\U{Y}$ & $\U{B-L}$  	\\    
			 \hline \vspace{-2mm} \\
			$q_{\rm L}$     			    							& $\bm{3}$		& $\bm{2}$	&	$1/6$ & $1/3$\vspace{1mm} \\
			$u_{\rm R}$     			    							& $\bm{3}$		& $\bm{1}$	&	$2/3$ & $1/3$\vspace{1mm}	\\
			$d_{\rm R}$     			    							& $\bm{3}$		& $\bm{1}$	&	$-1/3$ & $1/3$\vspace{1mm}\\
			$\ell_{\rm L}$     			    							& $\bm{1}$		& $\bm{2}$	&	$-1/2$ & $-1$\vspace{1mm}\\
			$e_{\rm R}$     			    							& $\bm{1}$		& $\bm{1}$	&	$-1$ & $-1$	\vspace{1mm} \\
			$\nu_{\rm R}$     			    							& $\bm{1}$		& $\bm{1}$	&	$0$ & $-1$ \vspace{1mm}\\
			$H$     			    							& $\bm{1}$		& $\bm{2}$	&	$1/2$ & $0$	\vspace{1mm}\\
			$\chi$     			    							& $\bm{1}$		& $\bm{1}$	&	$0$ & $2$\vspace{1mm}	\\
			\hline
		\end{tabular} 
		\caption{Fields and their quantum numbers in the minimal B-L-SM. The last two columns represent the weak and $B-L$ hypercharges, which we denote as $Y$ and $Y_{\rm B-L}$ throughout the text.}
		\label{tab:charges}  
	\end{center}
\end{table} 

\subsection{The scalar sector}

The scalar potential of the B-L-SM reads
\begin{equation}
\label{eq:potential}
V(H,\chi)= m^2 H^\dagger H + \mu^2 \chi^\ast \chi + \lambda_1 (H^\dagger H)^2 + \lambda_2 \(\chi^\ast \chi\)^2 + \lambda_3  \chi^\ast \chi H^\dagger H
\end{equation}
where $H$ and $\chi$ are the Higgs doublet and the complex SM-singlet, respectively, whose real-valued components can be cast as 
\begin{equation}
\begin{aligned}
H = \dfrac{1}{\sqrt{2}} 
\begin{pmatrix}
-i \(\omega_1 - i \omega_2 \) \\
v + (h + i z)
\end{pmatrix}\,,	
\qquad
\chi = \dfrac{1}{\sqrt{2}} \[ x + \(h^\prime + i z^\prime\) \]\,.	
\end{aligned}
\end{equation}	
While $v$ and $x$ are the vacuum expectation values (\vevs) describing the classical ground state configurations of the theory, $h$ and $h^\prime$ represent radial quantum fluctuations around the minimum of the potential. There are four Goldstone directions denoted as $\omega_1$, $\omega_2$, $z$ and $z^\prime$ which are absorbed into longitudinal modes of the $W$, $Z$ and $Z^\prime$ gauge bosons once spontaneous symmetry breaking (SSB) takes place. The scalar potential \eqref{eq:potential} is bounded from below (BFB) whenever the conditions \cite{Basso:2010jm}
\begin{equation}
4 \lambda_1 \lambda_2  -  \lambda_3^2 > 0 \quad , \quad \lambda_1 , \lambda_2>0 \quad . 
\label{eq:BFB}
\end{equation}
are satisfied and the electric charge conserving vacuum
\begin{equation}
\mean{H} = \dfrac{1}{\sqrt{2}} 
\begin{pmatrix}
0 \\
v 
\end{pmatrix}	
\qquad
\mean{\chi} = \dfrac{x}{\sqrt{2}}
\label{eq:vacuum}
\end{equation}
is stable. Resolving the tadpole equations with respect to the \vevs, one obtains
\begin{equation}
	v^2 = \tfrac{-\lambda_2 m^2 + \tfrac{\lambda_3}{2}\mu^2}{\lambda_1 \lambda_2 - \tfrac{1}{4}\lambda_3^2} > 0
	\qquad
	\text{and}
	\qquad
	x^2 = \tfrac{-\lambda_1 \mu^2 + \tfrac{\lambda_3}{2}m^2}{\lambda_1 \lambda_2 - \tfrac{1}{4}\lambda_3^2} > 0\,,
	\label{eq:extremum}
\end{equation}
which imply, together with the BFB conditions \eqref{eq:BFB}, that
\begin{equation}
\lambda_2 m^2 < \tfrac{\lambda_3}{2} \mu^2 
\qquad
\text{and}
\qquad
\lambda_1 \mu^2 < \tfrac{\lambda_3}{2} m^2 \,.
\label{eq:sols}
\end{equation}
While the sign of $\lambda_1$ and $\lambda_2$ is positive, the inequalities \eqref{eq:sols} put further constraints on the signs of $m^2$, $\mu^2$ and $\lambda_3$ according to Tab.~\ref{tab:signs}.
\begin{table}[htb!]
	\begin{center}
		\begin{tabular}{ccccc}
			& $\mu^2 > 0$ & $\mu^2 > 0$ & $\mu^2 < 0$ & $\mu^2 < 0$  	\\
			& $m^2 > 0$ & $m^2 < 0$ & $m^2 > 0$ & $m^2 < 0$  	\\        
						\hline \vspace{-2mm} \\ 
			$\lambda_3 < 0 $     			    							& \xmark		& \checkmark	&	\checkmark & \checkmark \vspace{1mm}	\\
			$\lambda_3 > 0$     			    							& \xmark		& \xmark	&	\xmark &  \checkmark \vspace{1mm} \\
		\end{tabular} 
		\caption{Signs of parameters in the potential (\ref{eq:potential}).
While the \checkmark\,symbol indicates the solutions of the tadpole conditions \eqref{eq:sols} which, together with the positively-definite scalar mass spectrum, correspond to a minimum of the scalar potential, \xmark\,indicates unstable configurations.}
		\label{tab:signs}  
	\end{center}
\end{table} 
We see that if $\lambda_3$ is positive a minimum in the scalar potential can emerge provided that both $\mu^2$ and $m^2$ are not simultaneously positive. However, in our studies we have considered the $\lambda_3 > 0$ solution in the last column of Tab.~\ref{tab:signs} where both the $\SU{2}{L}$ isodoublet and the complex singlet mass parameters are negative.

Taking the Hessian matrix and evaluating it in the vacuum \eqref{eq:vacuum} one obtains
\begin{equation}
\bm{M}^2 =
\begin{pmatrix}
4 \lambda_2 x^2 & \lambda_3 v x \\ 
\lambda_3 v x   & 4 \lambda_1 v^2 
\end{pmatrix}\,,
\label{eq:hess}
\end{equation}
which can be rotated to the mass eigenbasis as
\begin{equation}
	\bm{m}^2 = {O^\dagger}_{i}{}^{m} M_{mn}^2 O^{n}{}_{j} = 
	\begin{pmatrix}
	m_{h_1}^2 & 0 \\ 
	0   & m_{h_2}^2 
	\end{pmatrix}\,,
\end{equation}
where the eigenvalues are
\begin{equation}
m_{h_{1,2}}^2 = \lambda_1 v^2 + \lambda_2 x^2 \mp \sqrt{(\lambda_1 v^2 - \lambda_2 x^2)^2 + (\lambda_3 x v)^2}\,,
\label{eq:eigvals}
\end{equation}
and the orthogonal rotation matrix $\bm{O}$ reads
\begin{equation}
	\bm{O} = 
	\begin{pmatrix}
	\cos \alpha_h & -\sin \alpha_h \\
	\sin \alpha_h & \cos \alpha_h 
	\end{pmatrix}\,.
	\label{eq:rotmat}
\end{equation}
The physical basis vectors $h_1$ and $h_2$ can then be written in terms of the gauge eigenbasis ones $h$ and $h^\prime$ as follows:
\begin{equation}
	\begin{pmatrix}
	h_1 \\
	h_2 
	\end{pmatrix}
	=
	\bm{O}
	\begin{pmatrix}
	h \\
	h^\prime 
	\end{pmatrix}\,.
	\label{eq:trans}
\end{equation}

In this article, we consider scenarios where $\U{B-L}$ is broken above the EW-scale such that $x > v$. 
In the case of decoupling $v/x\ll 1$, the scalar masses and mixing angle become particularly simple,
\begin{equation}
\sin \alpha_h \approx \dfrac{1}{2}\dfrac{\lambda_3}{\lambda_2} \dfrac{v}{x} \qquad
m_{h_1}^2 \approx 2 \lambda_1 v^2 \qquad m_{h_2}^2 \approx 2 \lambda_2 x^2
\label{eq:simplify}
\end{equation}
which represent a good approximation for most of the phenomenologically consistent points 
in our numerical analysis discussed below.

\subsection{The gauge sector}

The gauge boson and Higgs kinetic terms in the B-L-SM Lagrangian read
\begin{equation}
\begin{aligned}
\mathcal{L}_{\U{}} =  \abs{D_\mu H}^2 + \abs{D_\mu \chi}^2 -\dfrac{1}{4} F_{\mu \nu} F^{\mu \nu} -\dfrac{1}{4} F^\prime_{\mu \nu} F^{\prime \mu \nu} -\dfrac{1}{2} \kappa F_{\mu \nu} F^{\prime \mu \nu}\,,
\end{aligned}
\label{eq:Lu1}
\end{equation}
where $F^{\mu \nu}$ and $F^{\prime \mu \nu}$ are the standard $\U{Y}$ and $\U{B-L}$ field strength tensors, respectively, 
\begin{equation}
	F_{\mu \nu} = \partial_\mu A_\nu - \partial_\nu A_\mu 
	\qquad
	\text{and}
	\qquad
	 F^\prime_{\mu \nu} = \partial_\mu A^\prime_\nu - \partial_\nu A^\prime_\mu\,.
	 \label{eq:Fmn}
\end{equation}
written in terms of the gauge fields $A_\mu$ and $A_\mu^\prime$, respectively.
The $\kappa$ parameter in Eq.~\eqref{eq:Lu1} represents the $\U{Y} \times \U{B-L}$ gauge kinetic mixing 
while the Abelian part of the covariant derivative reads
\begin{equation}
	D_\mu \supset i g_1 Y A_\mu + i g_1^\prime Y_{\rm B-L} A_\mu^\prime\,,
\end{equation}
with $g_1$ and $g_1^\prime$ being the $\U{Y}$ and $\U{B-L}$ the gauge couplings, respectively, 
whereas the $Y$ and $B-L$ charges are specified in Tab.~\ref{tab:charges}. 

\subsubsection{\it Kinetic-mixing}

In order to study the kinetic mixing effects on physical observables it is convenient to rewrite the gauge kinetic terms in the canonical form, i.e.
\begin{equation}
	F_{\mu \nu} F^{\mu \nu} + F^\prime_{\mu \nu} F^{\prime \mu \nu} + 2 \kappa F_{\mu \nu} F^{\prime \mu \nu} \to B_{\mu \nu} B^{\mu \nu} + B^\prime_{\mu \nu} B^{\prime \mu \nu}\,.
	\label{eq:AtoB}
\end{equation}
A generic orthogonal transformation in the field space does not eliminate the kinetic mixing term. So, in order to satisfy Eq.~\eqref{eq:AtoB} an extra non-orthogonal transformation should be imposed such that Eq.~\eqref{eq:AtoB} is realized. Taking $\kappa = \sin \alpha$, a suitable redefinition of fields $\{A_\mu,A_\mu^\prime\}$ into $\{B_\mu, B_\mu^\prime\}$ that eliminates $\kappa$-term according to Eq.~\eqref{eq:Lu1} can be cast as
\begin{equation}
	\begin{pmatrix}
	A_\mu \\
	A^\prime_\mu 
	\end{pmatrix}
	=
	\begin{pmatrix}
	1 & -\tan \alpha \\
	0 & \sec \alpha 
	\end{pmatrix}
	\begin{pmatrix}
	B_\mu \\
	B^\prime_\mu 
	\end{pmatrix}\,,
	\label{eq:trans-kappa}
\end{equation}
such that in the limit of no kinetic-mixing, $\alpha = 0$. Note that this transformation is generic and valid for any basis in the field space. The transformation (\ref{eq:trans-kappa}) results in a modification of the covariant derivative that acquires two additional terms encoding the details of the kinetic mixing, i.e.
\begin{equation}
D_\mu \supset \partial_\mu + i \(\g{Y}{} \; Y + \g{BY}{} \; Y_{B-L}\) B_\mu + i \(\g{B-L}{} \; Y_{B-L} + \g{YB}{} \; Y\) B_\mu^\prime\,,
\label{eq:newCov}
\end{equation}	
where the gauge couplings take the form
\begin{equation}
	\begin{cases}
	\g{Y}{} = g_1 \\
	\g{B-L}{} = g_1^\prime \sec \alpha \\
	\g{YB}{} = -g_1 \tan \alpha \\
	\g{BY}{} = 0
	\end{cases} \,,
	\label{eq:new-g-simp}
\end{equation}
which is the standard convention in the literature. The resulting mixing between the neutral gauge fields including $Z^\prime$ can be represented as follows
\begin{equation}
\begin{aligned}
\begin{pmatrix}
\gamma_\mu \\
Z_\mu \\
Z^\prime_\mu
\end{pmatrix}
=
\begin{pmatrix}
\cos \theta_W & \sin \theta_W & 0\\
-\sin \theta_W \cos \theta_W^\prime & \cos \theta_W \cos \theta_W^\prime & \sin \theta_W^\prime \\
\sin \theta_W \sin \theta_W^\prime & -\cos \theta_W^\prime \sin \theta_W^\prime & \cos \theta_W^\prime
\end{pmatrix}
\begin{pmatrix}
B_\mu \\
A^3_\mu \\
B^\prime_\mu
\end{pmatrix}
\end{aligned}
\label{eq:g-Z-Zp}
\end{equation}	
where $\theta_W$ is the weak mixing angle and $\theta^\prime_W$ is defined as
\begin{equation}
\sin(2 \theta^\prime_W) = \frac{2 \g{YB}{} \sqrt{g^2 + \g{Y}{2}}}{\sqrt{(\g{YB}{2} + 16 (\frac{x}{v})^2 \g{B-L}{2} - g^2 - \g{Y}{2})^2 + 4 \g{YB}{2} (g^2 + \g{Y}{2})} }\,,
\label{eq:theta-p-full}
\end{equation}
in terms of $g$ and $\g{Y}{}$ being the $\SU{2}{L}$ and $\U{Y}$ gauge couplings, respectively. In the physically relevant limit, $v/x \ll 1$, the above expression greatly simplifies leading to
\begin{equation}
	\sin \theta_W^\prime \approx \dfrac{1}{8
	} \dfrac{\g{YB}{}}{\g{B-L}{2}}\(\dfrac{v}{x}\)^2 \sqrt{g^2 + \g{Y}{2}} \,,
	\label{eq:theta-p}
\end{equation}
up to $(v/x)^3$ corrections. In the limit of no kinetic mixing, i.e. $\g{YB}{} \to 0$, there is no mixture of $Z^\prime$ and SM gauge bosons. 

Note, the kinetic mixing parameter $\theta_W^\prime$ has rather stringent constraints from $Z$ pole experiments both at the Large Electron-Positron Collider (LEP) \cite{ALEPH:2005ab} and the Stanford Linear Collider (SLC) \cite{Aaltonen:2010ws}, restricting its value to be smaller than $10^{-3}$ approximately, which we set as an upper bound in our numerical analysis. Expanding the kinetic terms $\abs{D_\mu H}^2 + \abs{D_\mu \chi}^2$ around the vacuum one can extract the following mass matrix for vector bosons
\begin{equation}
	m_V^2 =
	\dfrac{v^2}{4}
	\begin{pmatrix}
	g^2 \;\;&\;\; 0 \;\;&\;\; 0 \;\;&\;\; 0 \;\;&\;\; 0 \\
	0 \;\;&\;\; g^2 \;\;&\;\; 0 \;\;&\;\; 0 \;\;&\;\; 0 \\
	0 \;\;&\;\; 0 \;\;&\;\; g^2 \;\;&\;\; -g \g{Y}{} \;\;&\;\; -g \g{YB}{} \\
	0 \;\;&\;\; 0 \;\;&\;\; -g \g{Y}{} \;\;&\;\; \g{Y}{2} \;\;&\;\; \g{Y}{} \g{YB}{} \\
	0 \;\;&\;\; 0 \;\;&\;\; -g \g{YB}{} \;\;&\;\; \g{Y}{} \g{YB}{} \;\;&\;\; \g{YB}{2} + 16 \(\dfrac{x}{v}\)^2 \g{B-L}{2}
	\end{pmatrix}
\end{equation}
whose eigenvalues read
\begin{equation}
	m_A = 0 \, \text{,} \qquad m_W = \tfrac{1}{2} v g
\end{equation}
corresponding to physical photon and $W^\pm$ bosons as well as
\begin{equation}
m_{Z,Z^\prime}=\sqrt{g^2 + g^2_\ro{Y}} \cdot \frac{v}{2}  \sqrt{\frac{1}{2} \left( \frac{\g{YB}{2} + 16 (\frac{x}{v})^2 g^2_{\rm BL} }{g^2 + g^2_{\rm Y}} +1  \right) \mp \frac{\g{YB}{}}{\sin(2 \theta_W^\prime) \sqrt{g^2 + g^2_{\rm Y}}}}\,.
\label{eq:ZZp-mass}
\end{equation}
for two neutral massive vector bosons, with one of them, not necessarily the lightest, representing the SM-like $Z$ boson. It follows from LEP and SLC constraints on $\theta_W^\prime$, that Eq.~\eqref{eq:theta-p} also implies that either $\g{YB}{}$ or the ratio $\tfrac{v}{x}$ are small. In this limit, Eq.~\eqref{eq:ZZp-mass} simplifies to
\begin{equation}
	m_Z \approx \tfrac{1}{2} v \sqrt{g^2 + \g{Y}{2}} \qquad \text{and} \qquad m_{Z^\prime} \approx 2 \g{B-L}{} x\,,
	\label{eq:mZ}
\end{equation}
where the $m_{Z^\prime}$ depends only on the SM-singlet \vev\,$x$ and on the $\U{B-L}$ gauge coupling and will be attributed to a heavy $Z^\prime$ state, while the light $Z$-boson mass corresponds to its SM value.

\subsection{The Yukawa sector}

One of the key features of the B-L-SM is the presence of non-zero neutrino masses. In its minimal version, such masses are generated via a type-I seesaw mechanism. The Yukawa Lagrangian of the model reads
\begin{equation}
\begin{aligned}
\mathcal{L}_f = 
-Y_u^{ij} \overline{q_{\rm L i}} u_{\rm R j} \widetilde{H} 
-Y_d^{ij} \overline{q_{\rm L i}} d_{\rm R j} H
-Y_e^{ij} \overline{\ell_{\rm L i}} e_{\rm R j} H
- Y_\nu^{ij} \overline{\ell_{\rm L i}} \nu_{\rm R j} \widetilde{H}
	-\dfrac{1}{2} Y_\chi^{ij} \overline{\nu_{\rm R i}^c} \nu_{\rm R j} \chi + {\rm c.c.}
\end{aligned}
\label{eq:Yuk}
\end{equation}
Notice that Majorana neutrino mass terms of the form $M \overline{\nu_\ro{R}^c} \nu_\ro{R}$ would explicitly violate the $\U{B-L}$ symmetry and are therefore not present. In Eq.~\eqref{eq:Yuk}, $Y_u$, $Y_d$ and $Y_e$ are the $3 \times 3$ Yukawa matrices that reproduce the quark and charged lepton sector of the SM, while $Y_\nu$ and $Y_\chi$ are the new Yukawa matrices responsible for the generation of neutrino masses and mixing. In particular, one can write
\begin{equation}
	\bm{m}_{\nu_l}^\ro{Type-I} = \dfrac{1}{\sqrt{2}}\dfrac{v^2}{x} \bm{Y}_\nu^\top \bm{Y}^{-1}_\chi \bm{Y}_\nu\,,
\end{equation}
for light $\nu_l$ neutrino masses, whereas the heavy $\nu_h$ ones are given by
\begin{equation}
	\bm{m}_{\nu_h}^\ro{Type-I} \approx \dfrac{1}{\sqrt{2}} \bm{Y}_\chi x\,,
\end{equation} 
where we have assumed a flavour diagonal basis. Note that the smallness of light neutrino masses imply that either the $x$ VEV is very large or (if we fix it to be at the $\mathcal{O}\(\ro{TeV}\)$ scale and $\bm{Y}_\chi \sim \mathcal{O}\(1\)$) the corresponding Yukawa coupling should be tiny, $\bm{Y}_\nu < 10^{-6}$. It is clear that the low scale character of the type-I seesaw mechanism in the minimal B-L-SM is \textit{faked} by small Yukawa couplings to the Higgs boson. A more elegant description was proposed in Ref.~\cite{Khalil:2010iu} where small SM neutrino masses naturally result from an inverse seesaw mechanism. In this work, however, we will not study the neutrino sector and thus, for an improved efficiency of our numerical analysis of $Z^\prime$ observables, it will be sufficient to fix the Yukawa couplings to $\bm{Y}_\chi = 10^{-1}$ and $\bm{Y}_\nu = 10^{-7}$ values such that the three lightest neutrinos lie in the sub-eV domain.

\section{Parameter space studies}
\label{sec:parameter-space-studies}

To assess the phenomenological viability of the minimal B-L-SM, we have developed a scanning routine that sequentially calls publicly available software tools in order to numerically evaluate physical observables and confront them against experimental data. Analytical expressions for such observables are calculated in \texttt{SARAH 4.13.0} \cite{Staub:2008uz,Staub:2013tta} and then imported to \texttt{SPheno 4.0.3} \cite{Porod:2003um,Porod:2011nf}, which is a spectrum generator where masses and mixing angles, EW precision observables, the muon anomalous magnetic moment as well as a number of decay widths and branching fractions are numerically evaluated. Besides, various theoretical constraints such as the positivity of the one-loop mass spectrum and unitarity are taken into account. As a first step, our scanning routine randomly samples parameter space points according to the ranges in Tab.~\ref{tab:scan}.
\begin{table}[htb!]
\begin{center}
		\begin{tabular}{ccccc}
			$\lambda_{1}$ & $\lambda_{2,3}$ & $g_\ro{B-L}$ & $g_\ro{YB}$ & $x~{\rm [TeV]}$  
			\\       
						\hline \vspace{-2mm} \\ 
			$\[10^{-2},\; 10^{0.5}
			\]$ 			    							& $\[10^{-8},\; 10
			\]$ 			    							& $\[10^{-8},\; \sqrt{4\pi}
			\]$		& $\[10^{-8},\; \sqrt{4\pi}
			\]$	&	$\[0.5,\; 20.5
			\]$ 	\\
		\end{tabular}  
		\caption{Parameter scan ranges used in our analysis. Note that the value of $\lambda_1$ is mostly constrained by the tree-level Higgs boson mass given in Eq.~\eqref{eq:simplify}. 
		}
		\label{tab:scan}
\end{center}
\end{table}
As can be seen from Eq.~\eqref{eq:simplify}, $\lambda_1$ varies in a rather narrow domain in comparison to $\lambda_{2,3}$ in order to comply with the experimental data on the SM Higgs mass (in the limit of large singlet VEV). In particular, provided that \texttt{SPheno} computes the SM Higgs boson mass at two-loop order, the tree-level quantity $v \sqrt{2 \lambda_1}$ must not be too far from $\mathcal{O}\(125~\ro{GeV}\)$ for most of the valid points. In fact, we have verified that valid points typically require $\lambda_1 \sim \mathcal{O}\(0.12 - 0.14\)$, with a few cases where quantum corrections are somewhat larger. For the singlet VEV $x$, we scan over all its potentially phenomenologically interesting ranges, covering both large and small $Z’$ masses and both heavy 
and light second Higgs boson. In particular, we aim at exploring a specific domain in the parameter space where a heavy $Z’$ is still compatible with a relatively light $h_2$. 
As we will discuss below, our results demonstrate that a $Z’$ boson with mass up to 10 TeV is still compatible with sub-TeV second Higgs state in the considered BL-SM.

\subsection{Phenomenological constraints}

\begin{figure}[!htb]
\centering
\includegraphics[scale=0.29]{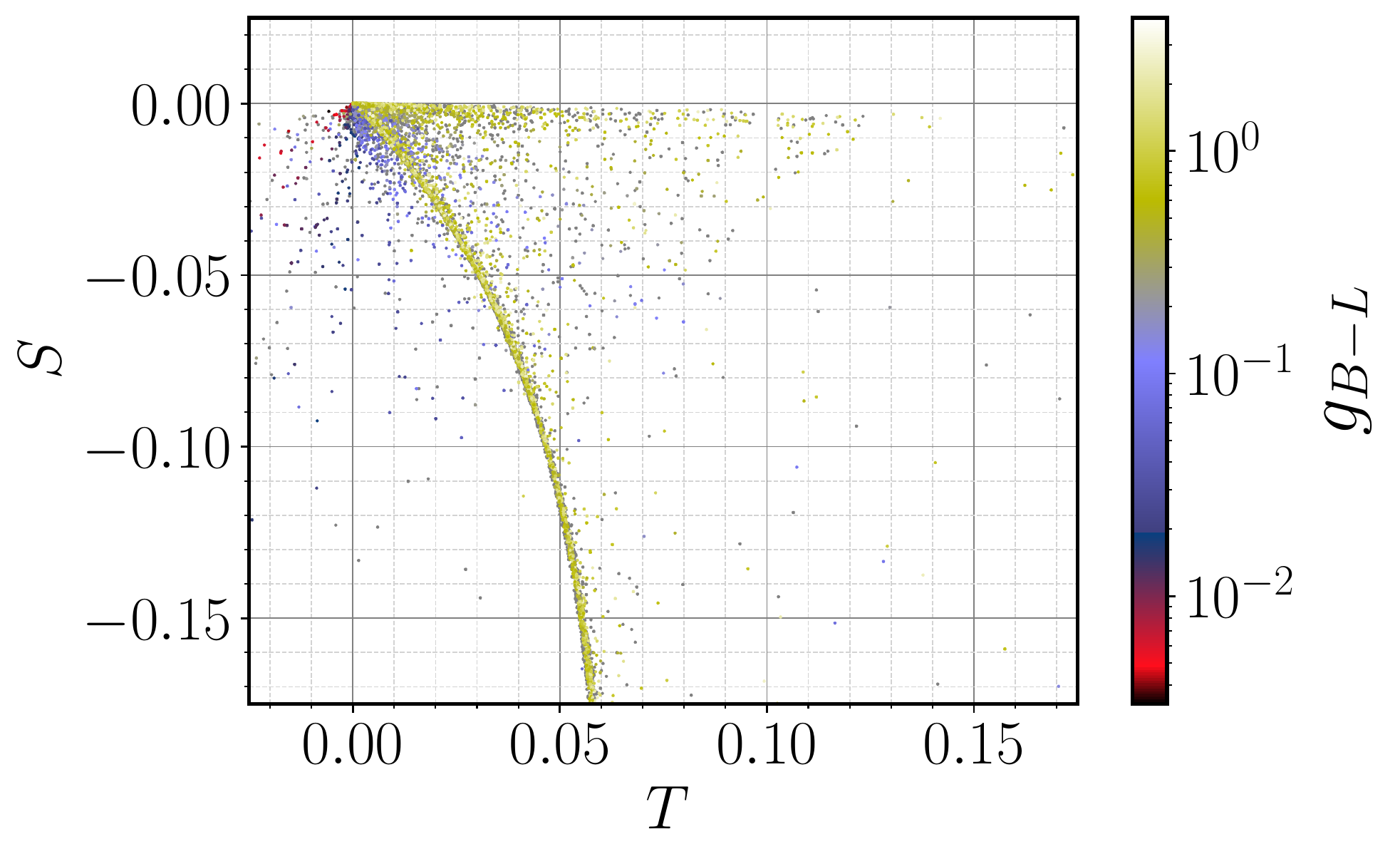}
\includegraphics[scale=0.29]{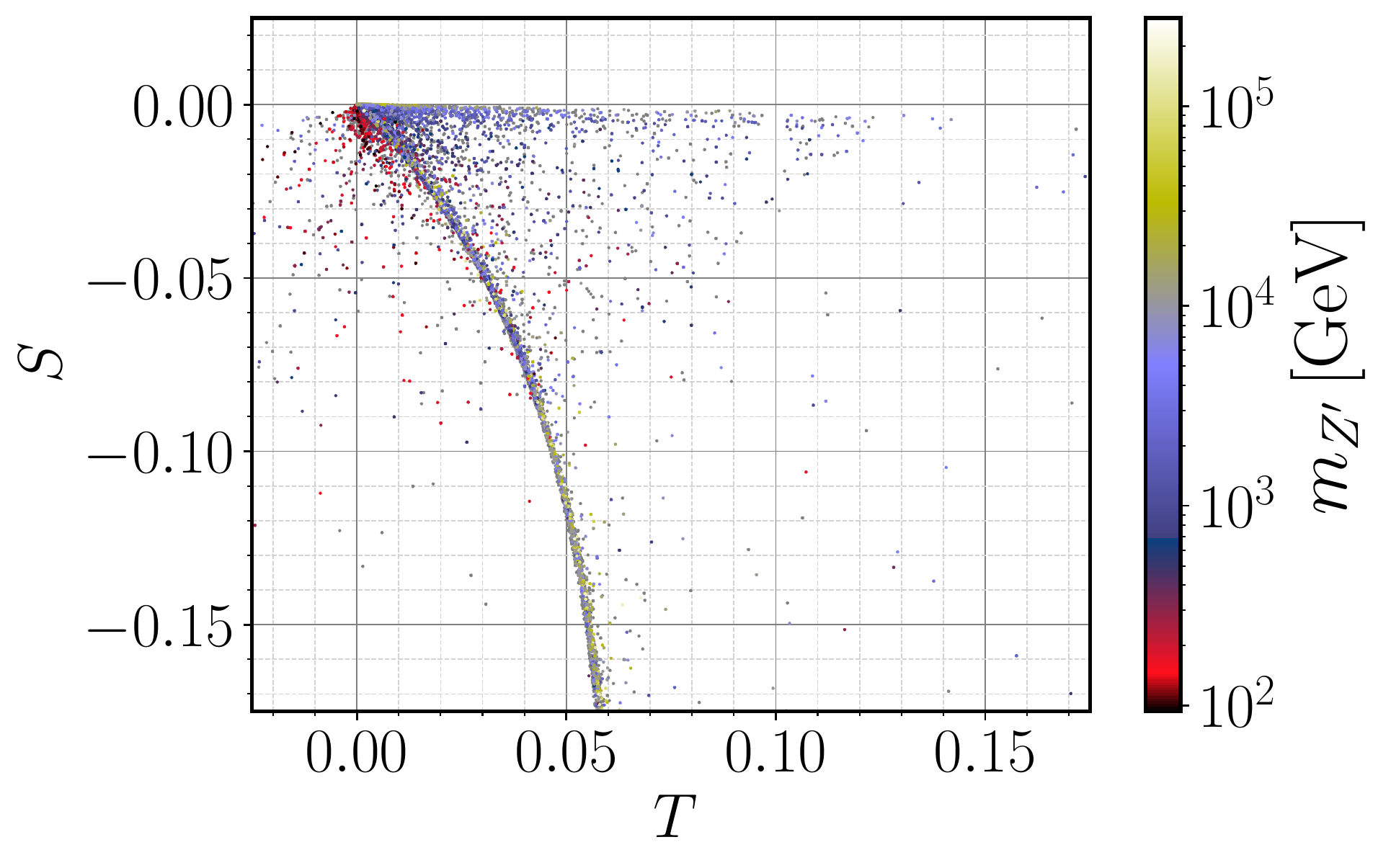}
\includegraphics[scale=0.29]{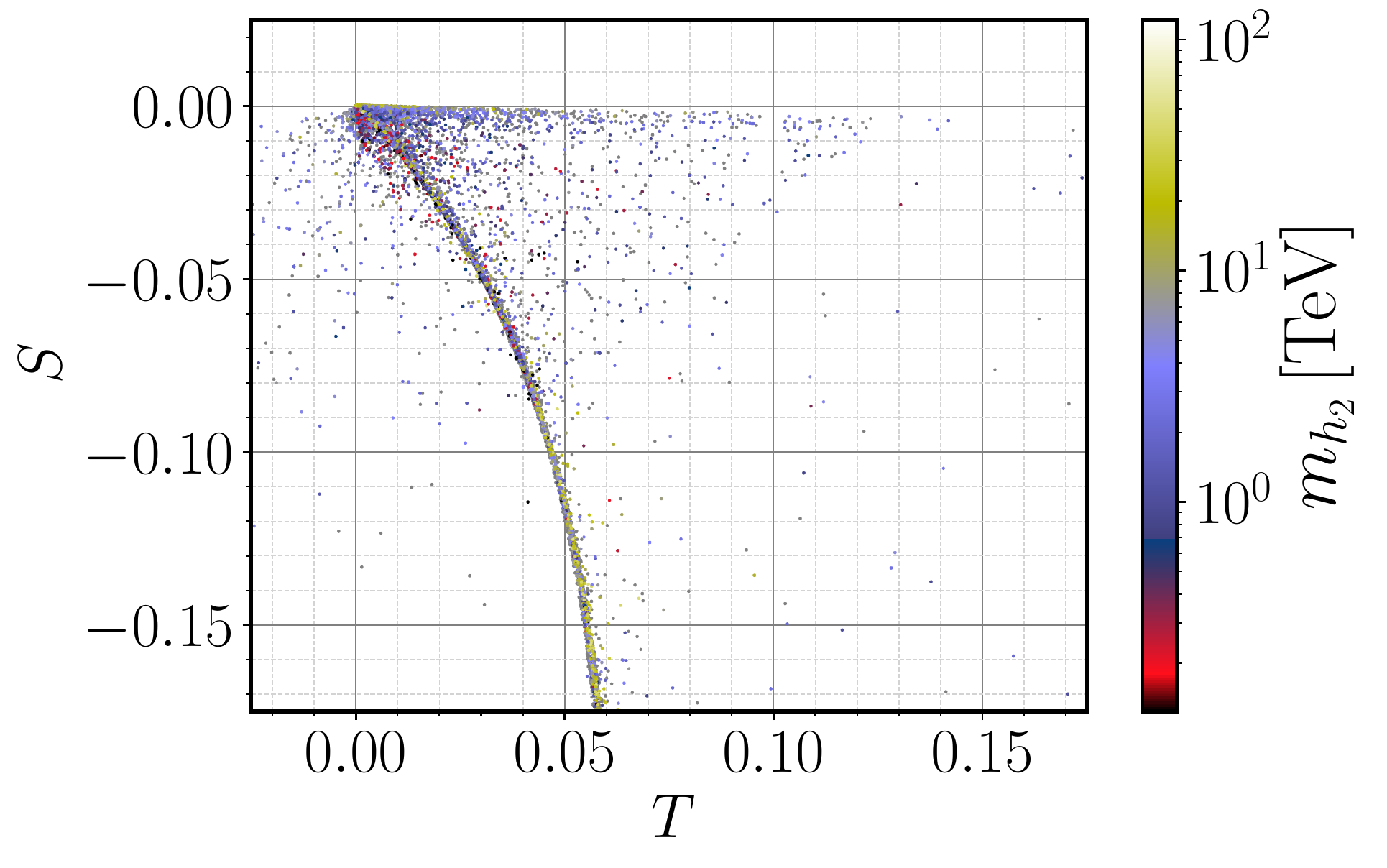}
\includegraphics[scale=0.29]{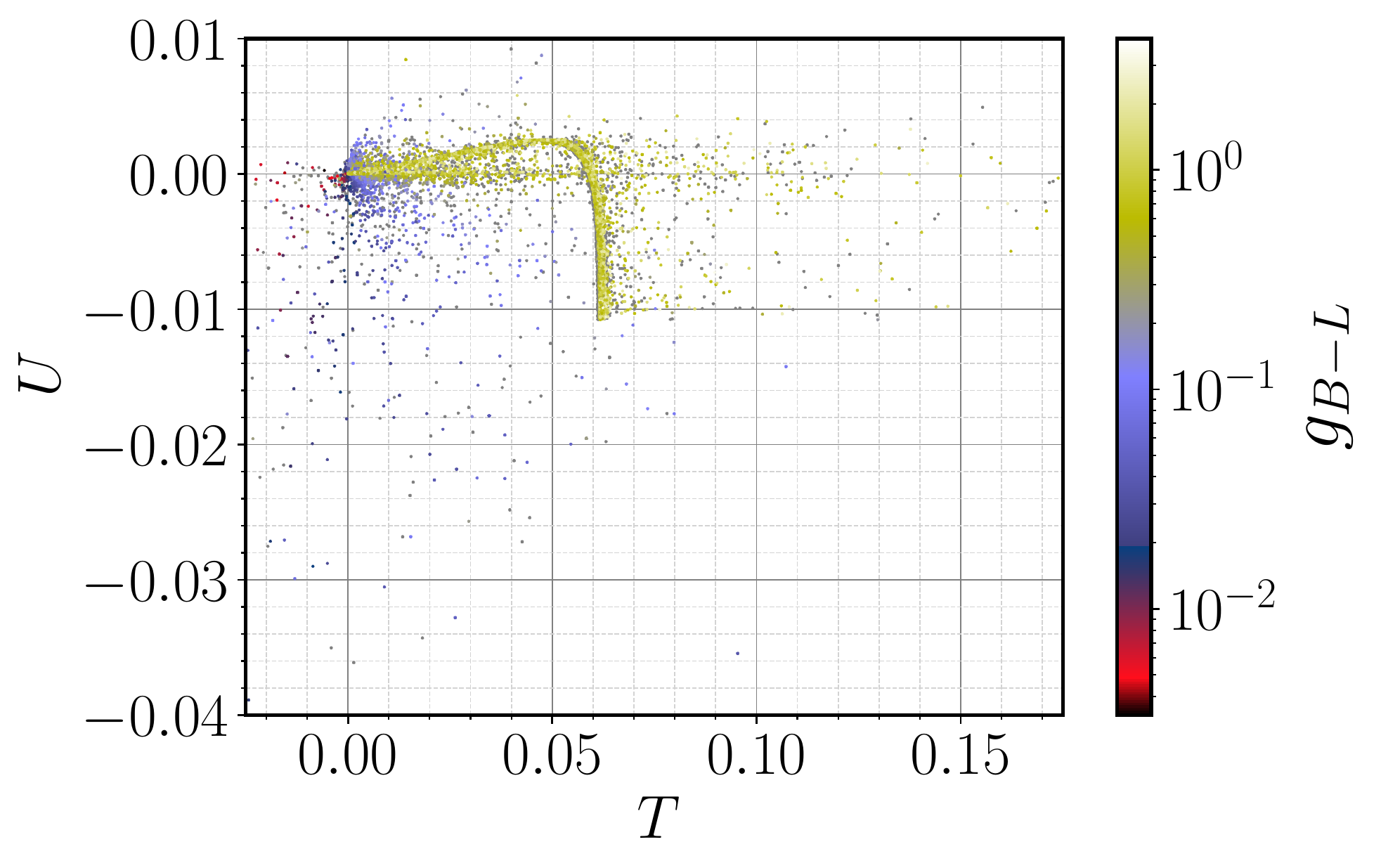}
\includegraphics[scale=0.29]{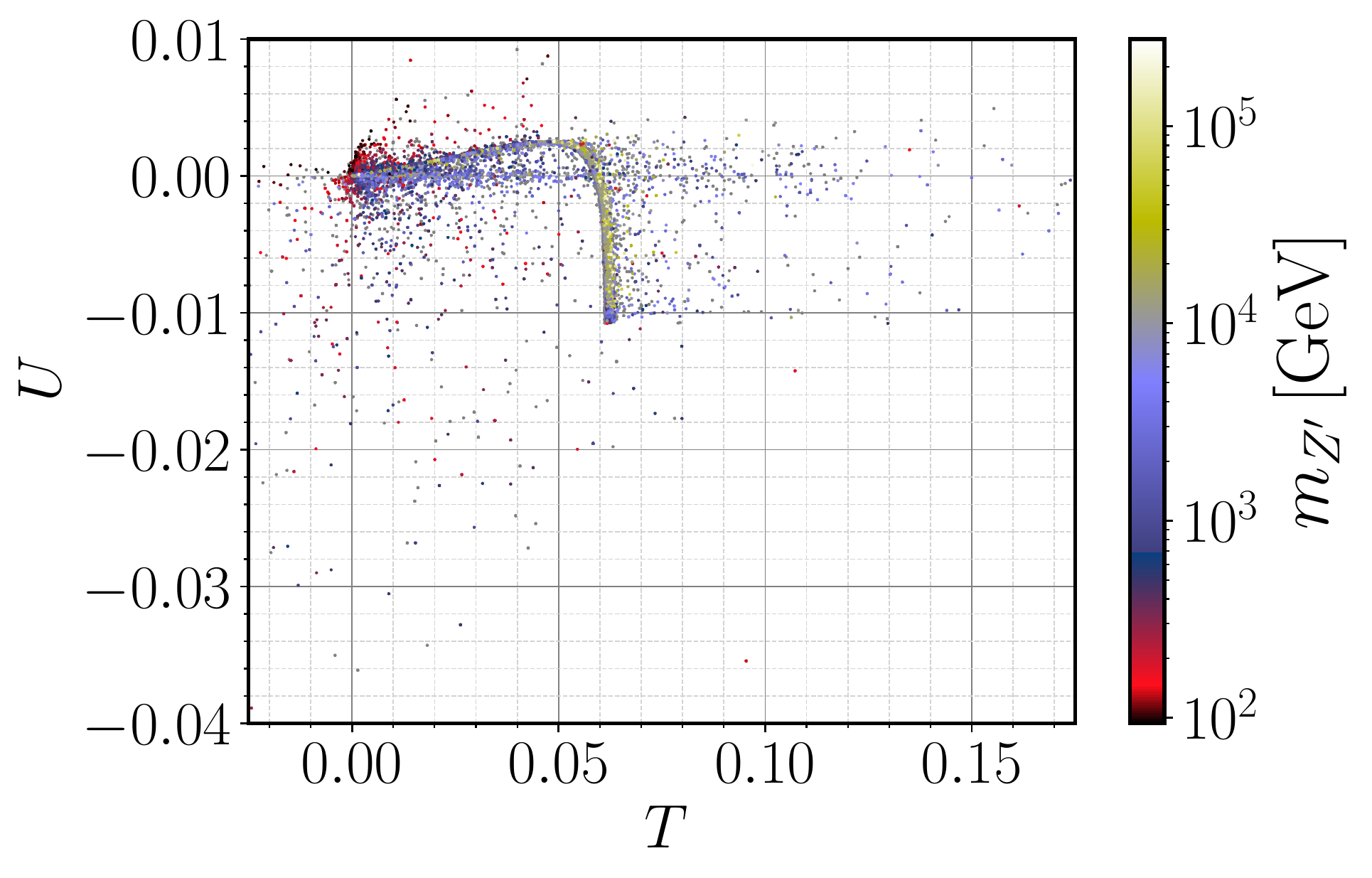}
\includegraphics[scale=0.29]{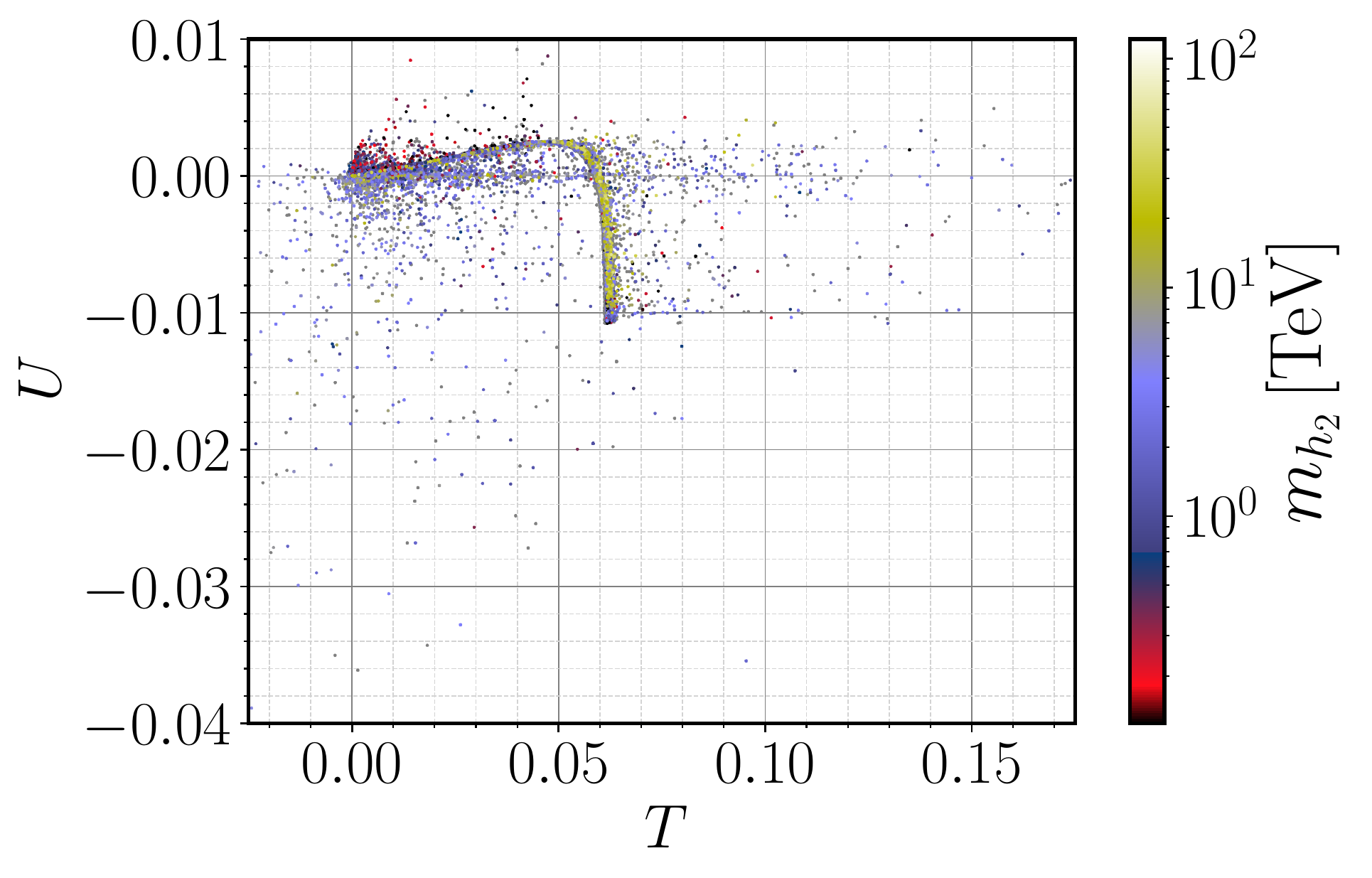}
\caption{Scatter plots for the EW oblique corrections in the $ST$ (upper row) and $TU$ (lower row) planes. In the color scales, 
their correlations with the $\U{B-L}$ gauge coupling $g_\ro{B-L}$, $Z^\prime$ mass, $m_{Z^\prime}$, and second Higgs mass, $m_{h_2}$, are shown in left, 
middle and right panels, respectively. The points shown here are passed the unitarity constraints in \texttt{SPheno 4.0.3} \cite{Porod:2003um,Porod:2011nf} as well as the 
Higgs phenomenology constraints in \texttt{HiggsBounds 4.3.1} \cite{Bechtle:2013wla} and \texttt{HiggsSignals 1.4.0} \cite{Bechtle:2013xfa}.}
\label{fig:STU-gBL_Zp_HP}
\end{figure}

In the B-L-SM, new physics (NP) contributions to $a_\mu$, denoted as $\Delta a_\mu^\ro{NP}$ in what follows, can emerge from the diagrams containing $Z^\prime$ or $h_2$ propagators. In this article, we study whether the muon anomalous magnetic moment can be at least partially explained in the model under consideration. Each parameter space point generated with our routine undergoes a sequence of tests before getting accepted. The very first layer of phenomenological checks is done by \texttt{SPheno} which promptly rejects any scenario with tachyonic scalar masses. If the positivity of squared scalar spectrum is assured, then \texttt{SPheno} verifies if unitary constraints are also fulfilled. For details see the pioneering work in \cite{Lee:1977eg} or the discussion in \cite{Coimbra:2013qq}. The presence of new bosons in the theory can induce large deviations in EW precision observables. Typically, the most stringent constraints emerge from the oblique $S,T,U$ parameters \cite{Kennedy:1988sn,Peskin:1990zt,Maksymyk:1993zm}, which are calculated by \texttt{SPheno}. In Fig.~\ref{fig:STU-gBL_Zp_HP},
we present the results for the EW oblique corrections in the $ST$ (upper row) and $TU$ (lower row) planes, together with their correlations with respect to 
the $\U{B-L}$ gauge coupling $g_\ro{B-L}$ (left), $Z^\prime$ mass, $m_{Z^\prime}$ (middle), and second Higgs mass, $m_{h_2}$ (right) shown in the color scale.

\begin{figure}[!htb]
\centering
\includegraphics[scale=0.29]{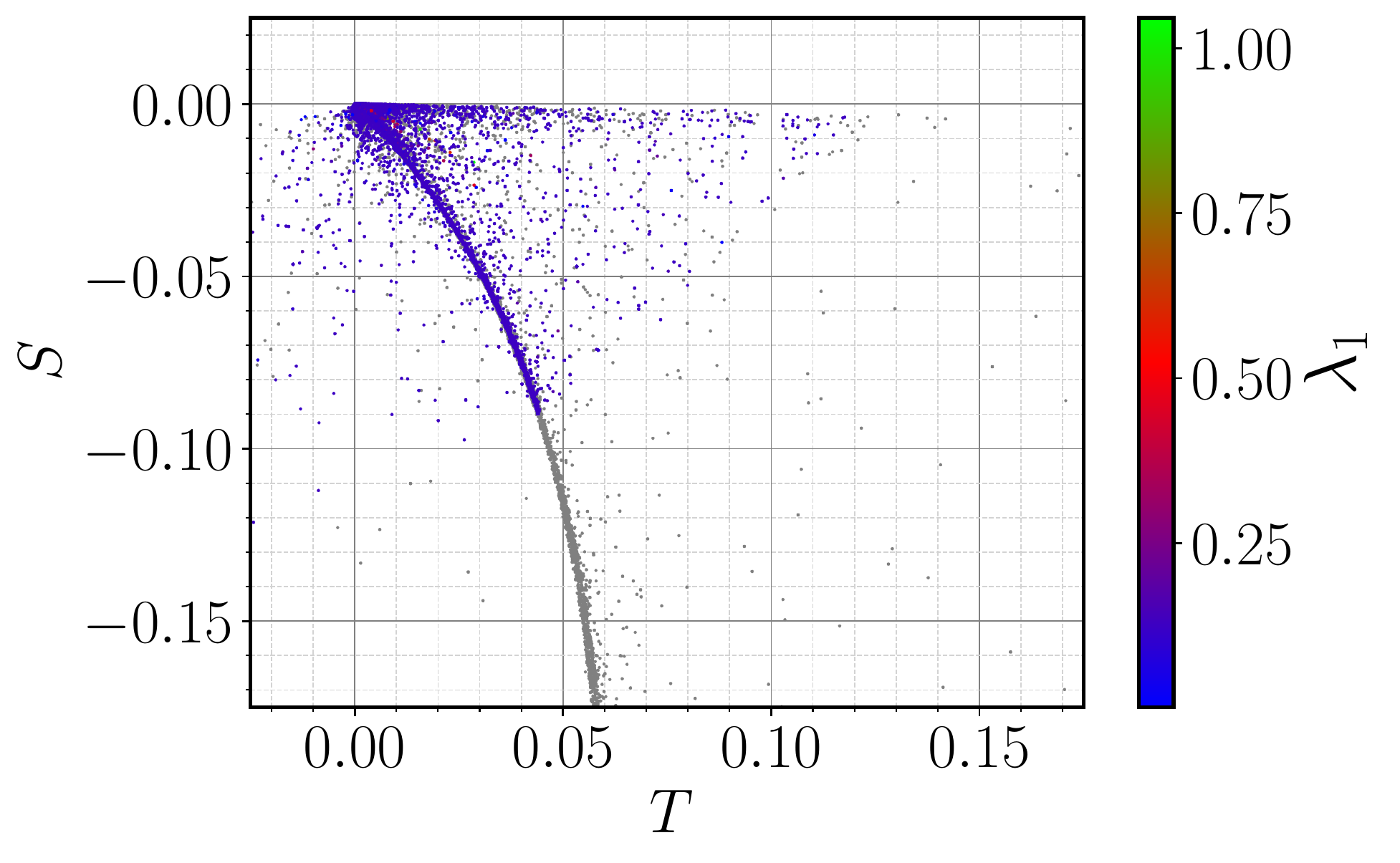}
\includegraphics[scale=0.29]{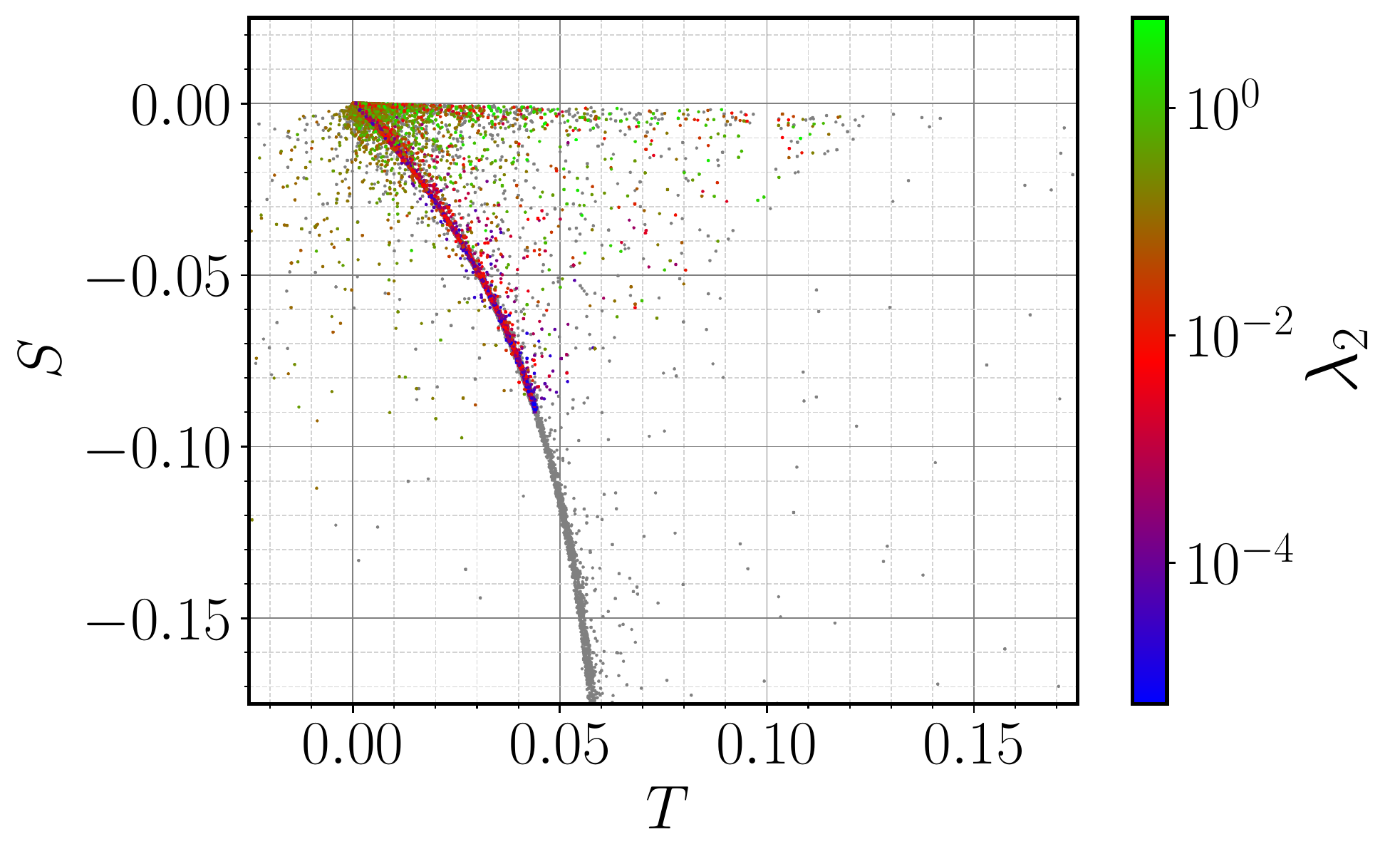}
\includegraphics[scale=0.29]{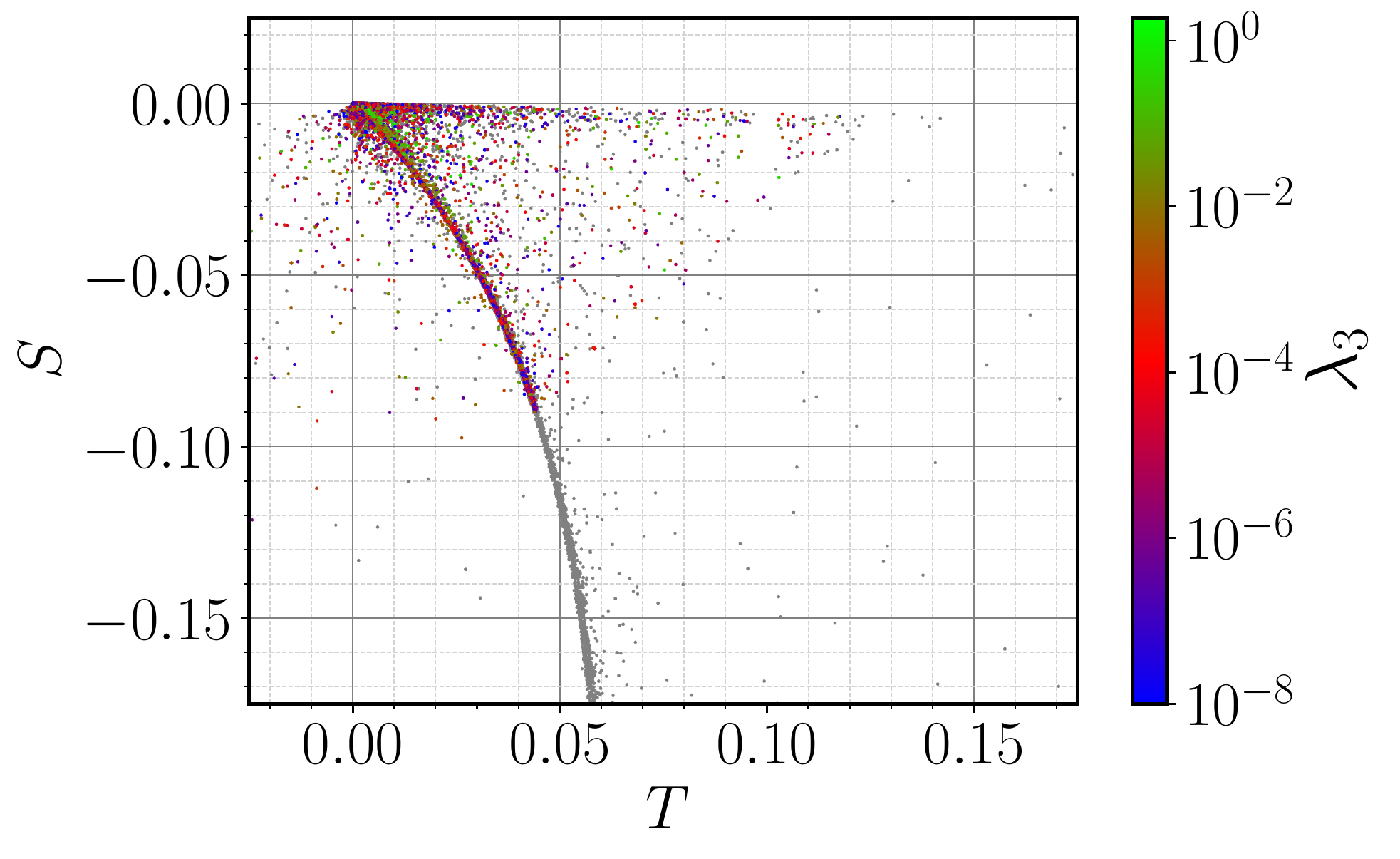}
\includegraphics[scale=0.29]{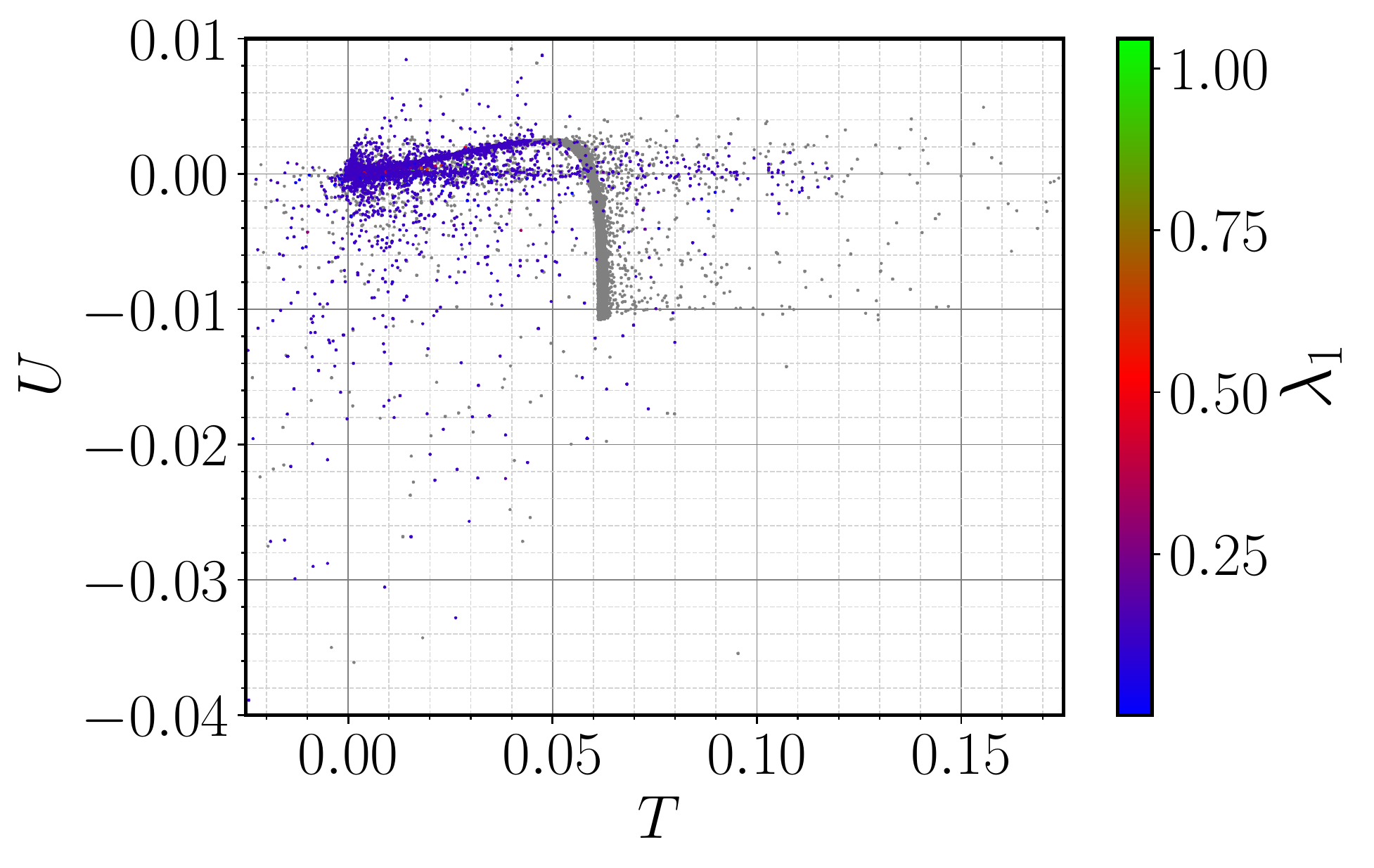}
\includegraphics[scale=0.29]{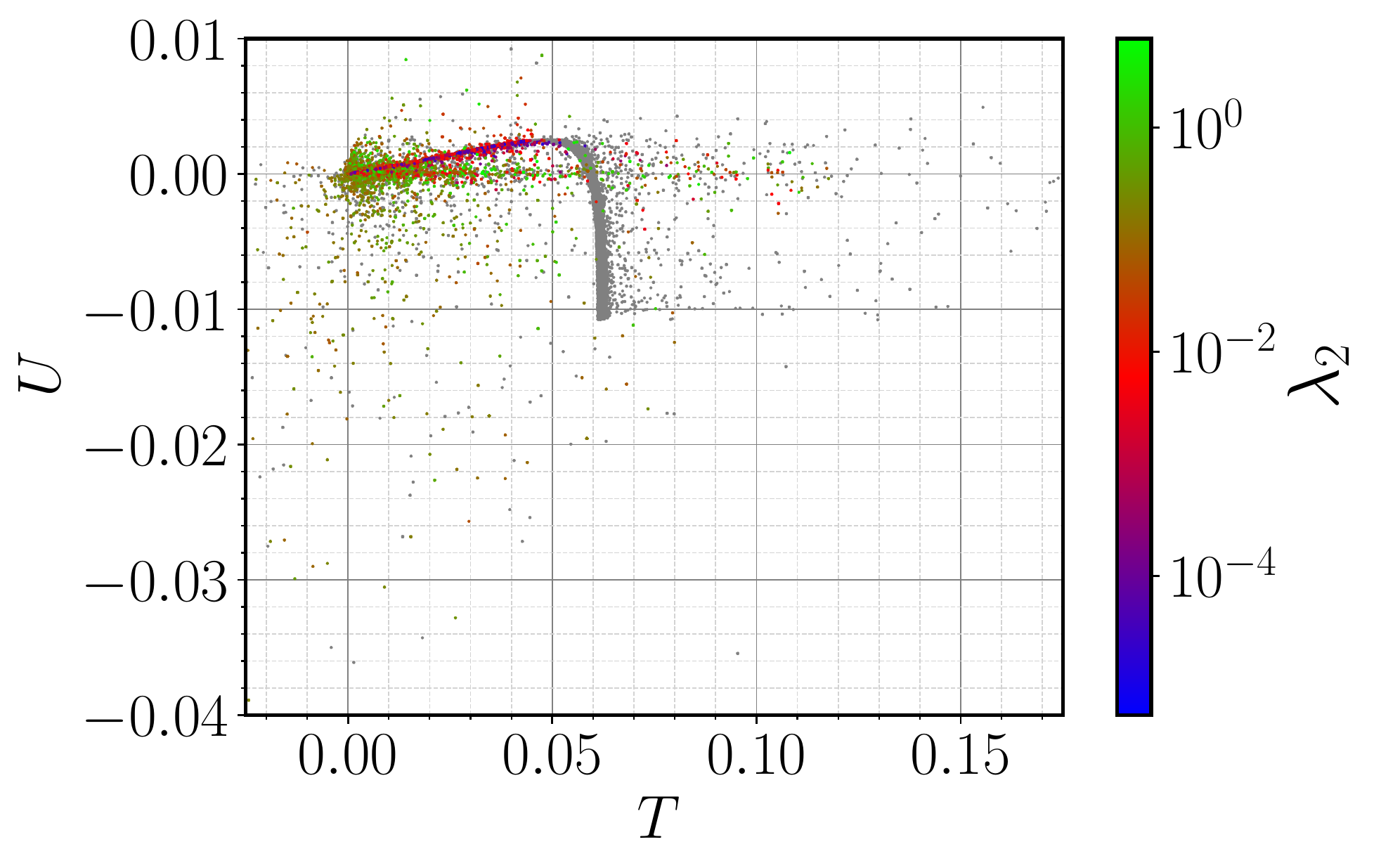}
\includegraphics[scale=0.29]{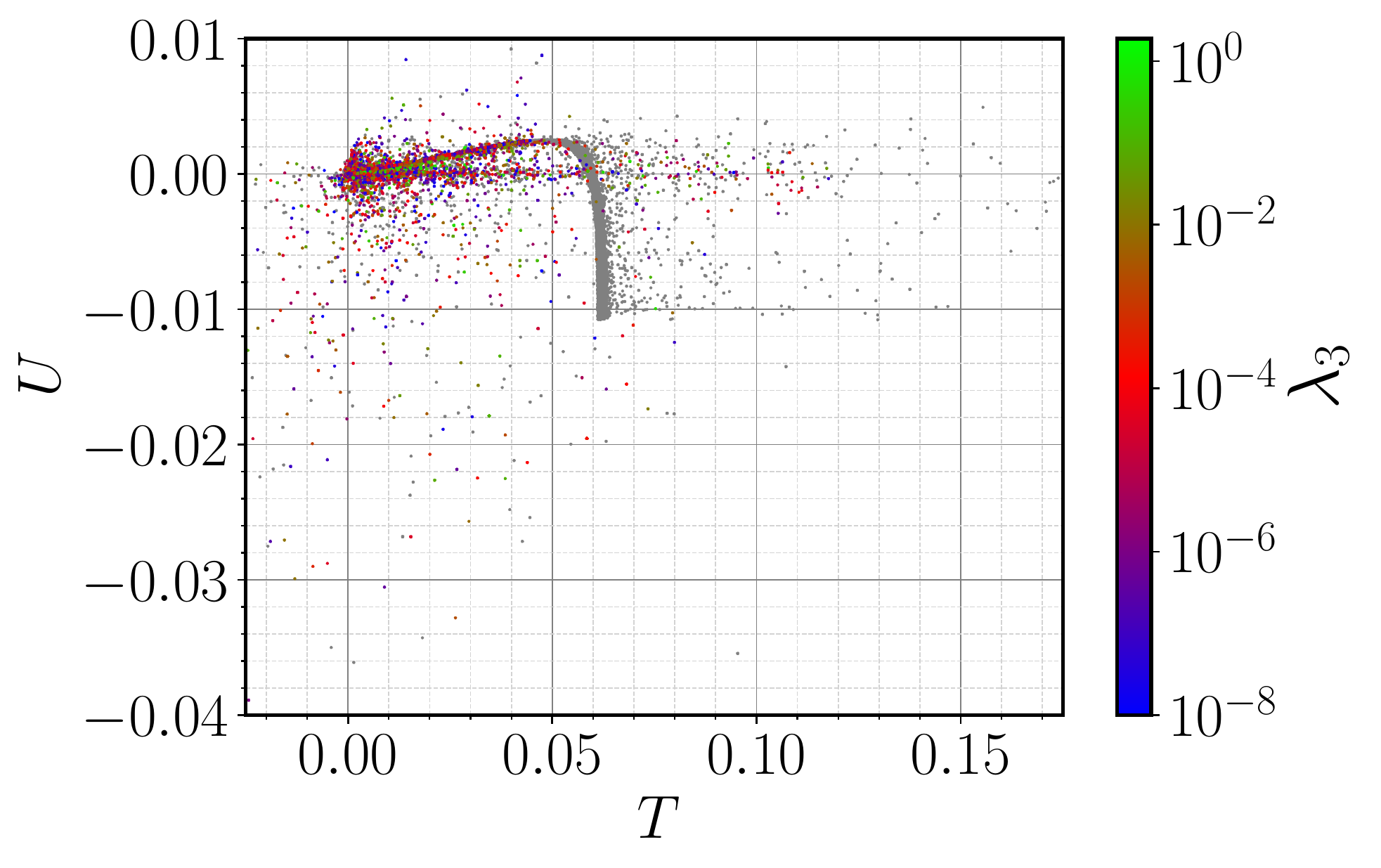}
\caption{Scatter plots for the EW oblique corrections in the $ST$ (upper row) and $TU$ (lower row) planes versus the scalar quartic 
couplings $\lambda_{1,2,3}$ shown in the color scale.
Accepted points lying within a $95\%$ C.L.~ellipsoid of the best fit point, and hence satisfying the EW precision constraints, 
are given in color whereas grey points are excluded. Other constraints are the same as in Fig.~\ref{fig:STU-gBL_Zp_HP}.  }
\label{fig:STU-lambda-excl}
\end{figure}

Current precision measurements \cite{Tanabashi:2018oca} provide the allowed regions 
\begin{equation}
	S = 0.02 \pm 0.10\,, \qquad T = 0.07 \pm 0.12\,, \qquad U = 0.00 \pm 0.09
	\label{eq:oblique}
\end{equation}
where $S$-$T$ are $92\%$ correlated, while $S$-$U$ and $T$-$U$ are $-66\%$ and $-86\%$ anti-correlated, respectively. We compare our results with the EW fit in Eq.~\eqref{eq:oblique} and require consistency with the best fit point within a $95\%$ C.L.~ellipsoid (see Ref.~\cite{Costa:2014qga} for further details about this method). We show in Fig.~\ref{fig:STU-lambda-excl} our results in the $ST$ (upper row) and $TU$ (lower row) planes where coloured points are consistent with EW precision observables at $95\%$ C.L.~whereas grey ones lie outside the corresponding ellipsoid of the best fit point and, thus, are excluded in our analysis. The color scales show correlations with the scalar quartic couplings $\lambda_{1,2,3}$.

The B-L-SM predicts a new visible scalar, which we denote as $h_2$, in addition to a SM-like $125~\ro{GeV}$ Higgs boson, $h_1$. Thus, in a second layer of phenomenological tests in both 
Figs.~\ref{fig:STU-gBL_Zp_HP} and \ref{fig:STU-lambda-excl}, we implemented also the collider bounds on the Higgs sector. In particular, we use \texttt{HiggsBounds 4.3.1} \cite{Bechtle:2013wla} to apply $95\%$ C.L.~exclusion limits on a new scalar particle, $h_2$, and \texttt{HiggsSignals 1.4.0} \cite{Bechtle:2013xfa} to check for consistency with the observed Higgs boson taking into account all known Higgs signal data. For the latter, we have accepted points whose fit to the data replicates the observed signal at $95\%$ C.L.~while the measured value for its mass, $m_{h_1} = 125.10 \pm 0.14~\ro{GeV}$ \cite{Tanabashi:2018oca}, is reproduced within a $3\sigma$ uncertainty. The required input data for \texttt{HiggsBounds/HiggsSignals} are generated by the \texttt{SPheno} output in the format of a SUSY Les Houches Accord (SLHA) \cite{Skands:2003cj} file. In particular, it provides scalar masses, total decay widths, Higgs decay branching ratios as well as the SM-normalized effective Higgs couplings to fermions and bosons squared (that are needed for analysis of the Higgs boson production cross sections). For details about this calculation, see Ref.~\cite{Bechtle:2013wla}.

As the third layer of phenomenological tests, in this work we have studied the viability of the surviving scenarios from the perspective of direct collider searches for a new $Z^\prime$ gauge boson. We have used \texttt{MadGraph5\_aMC@NLO 2.6.2} \cite{Alwall:2014hca} to compute the $Z^\prime$ Drell-Yan production cross section and subsequent decay into the first- and second-generation leptons, i.e.~$ \sigma\(pp \to Z^\prime\) \times B\(Z^\prime \to \ell \ell\)$ with $\ell = e,\; \mu$, and then compared our results to the most recent ATLAS exclusion bounds from the LHC runs at the center-of-mass energy $\sqrt{s} = 13~\ro{TeV}$ and integrated luminosity of 139 fb$^{-1}$ \cite{Aad:2019fac}. The \texttt{SPheno} SLHA output files were used as parameter cards for \texttt{MadGraph5\_aMC@NLO}, where the information required to calculate $ \sigma\(pp \to Z^\prime\) \times B\(Z^\prime \to \ell \ell\)$, such as the $Z^\prime$ boson mass, its total width and decay branching ratios into lepton pairs, is provided. In accordance with the experimental analysis, we have imposed a transverse momentum cut of 30 GeV for both final-state leptons while their pseudorapidities were limited to $|\eta| < 2.5$. An analogical analysis by the CMS Collaboration \cite{Sirunyan:2018exx} relies on a more complicated set of kinematic variables. So in the current work, for simplicity, we have only considered the ATLAS bound on $ \sigma\(pp \to Z^\prime\) \times B\(Z^\prime \to \ell \ell\)$ that is sufficient for our purposes.

An important and rather restrictive constraint that needs to be taken into account results from LEP limits on four-fermion contact interactions \cite{Alcaraz:2006mx,Freitas:2014pua}. In particular, we see from Tab.~3.13 of \cite{Alcaraz:2006mx} that, for the B-L-SM, this translates into the $95\%~\mathrm{C.L.}$ upper bounds on the $\g{L,R}{\ell \ell Z'}$ couplings
\begin{equation}
	\g{L}{\ell \ell Z'} < 0.221238\;\Big( \frac{m_{Z'}}{\rm TeV} \Big) \qquad \g{R}{\ell \ell Z'} < 0.274518\;\Big( \frac{m_{Z'}}{\rm TeV} \Big)\,.
	\label{eq:contact}
\end{equation}
This also poses upper limits on the $\U{B-L}$ and kinetic-mixing gauge couplings $\g{B-L}{}$ and $\g{YB}{}$, respectively, which are related to $\g{L,R}{\ell \ell Z'}$ via Eq.~\eqref{eq:gllZ}.

\subsection{Discussion of numerical results}
\label{sec:discuss}

Let us now discuss the phenomenological properties of the B-L-SM model. First, we focus on the current collider constraints and study their impact on both the scalar and gauge sectors.
\begin{figure}[!htb]
	\centering
	\includegraphics[scale=0.37]{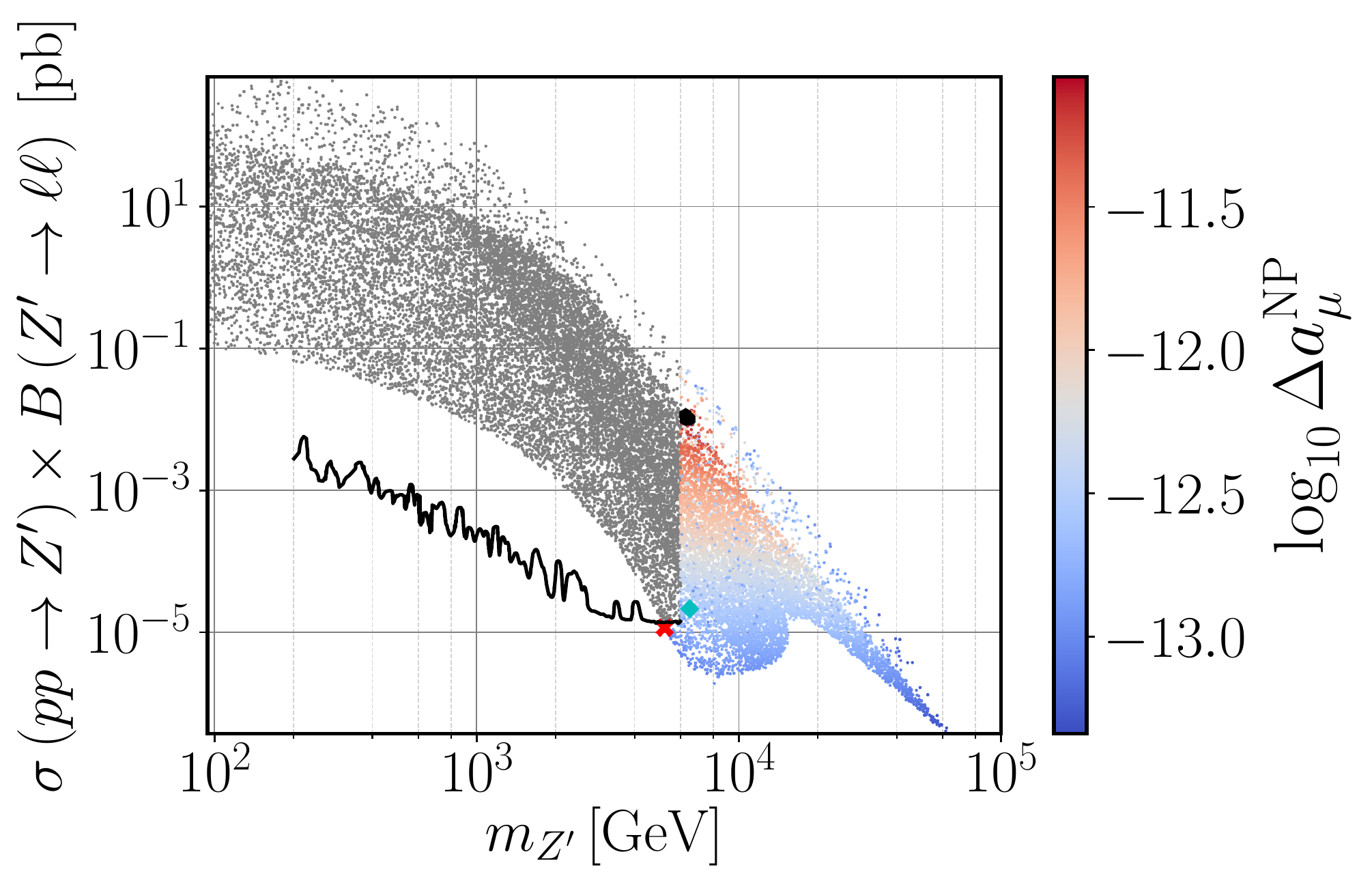}
	\includegraphics[scale=0.37]{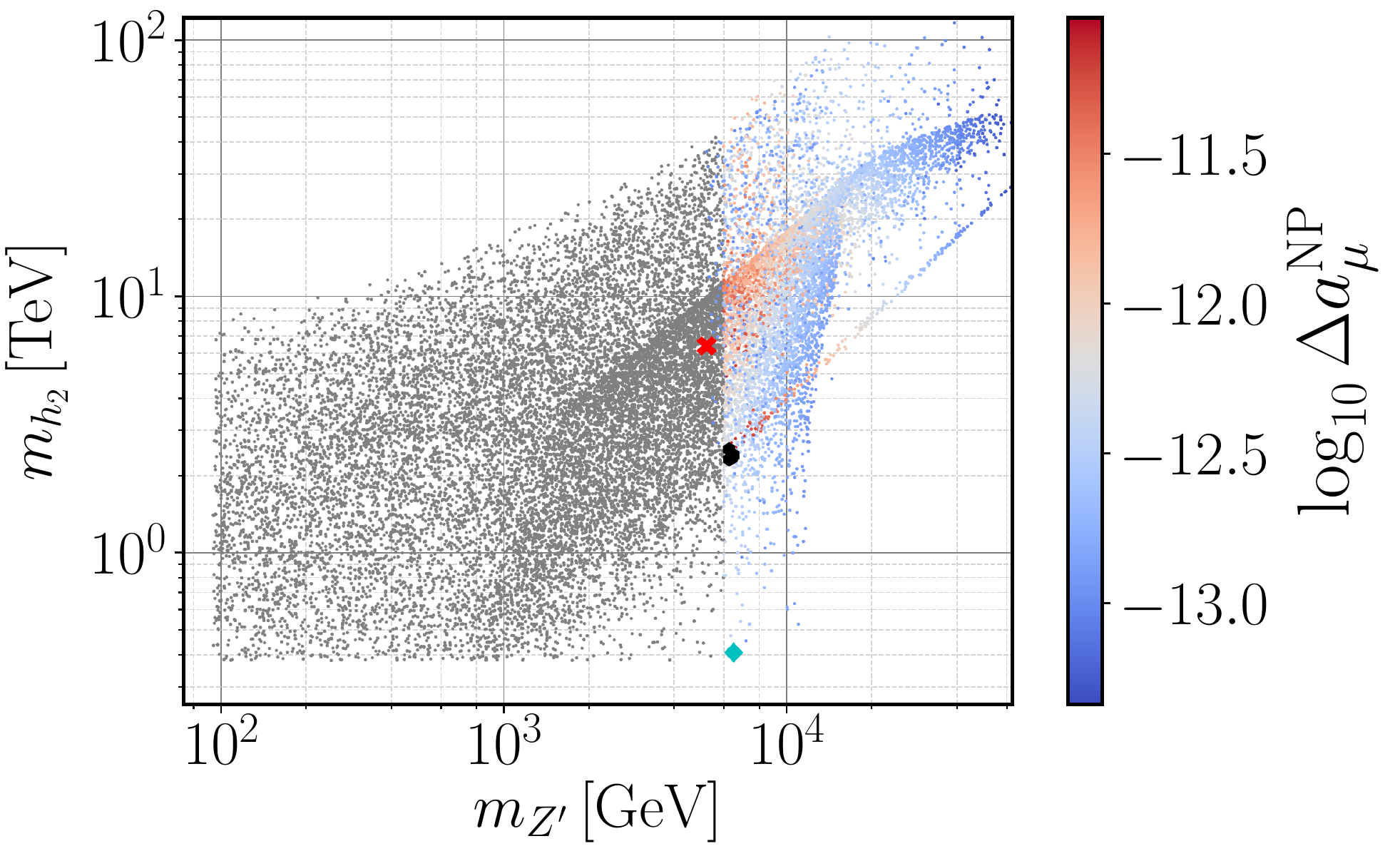}
	\caption{Scatter plots showing the $Z^\prime$ Drell-Yan production cross section times the decay branching ratio into a pair of electrons 
	and muons (left panel) and the new scalar mass $m_{h_2}$ (right panel) as functions of $m_{Z^\prime}$ and the new physics (NP) contributions 
	to the muon $\Delta a_\mu$ anomaly. The solid line represents the current ATLAS expected limit on the production cross section times branching 
	ratio into a pair of leptons at $95\%$ C.L.~ taken from Ref.~\cite{Aad:2019fac}.  Coloured points have survived all theoretical and experimental 
	constraints while grey points are excluded by direct $Z^\prime$ searches at the LHC. The six highlighted points in both panels denote the benchmark 
	scenarios described in Tab.~\ref{tab:bench}. These are represented by the red cross for the lightest $Z^\prime$ scenario (first row), cyan diamond 
	for the lightest $h_2$ scenario (second row) and the black dots (last four rows).}
	\label{fig:Plots1}
\end{figure}	

We show in Fig.~\ref{fig:Plots1} the scenarios generated in our parameter space scan (for the input parameter ranges, see Tab.~\ref{tab:scan}) that have passed all theoretical constraints such as boundedness from below, unitarity and EW precision tests, are compatible with the SM Higgs data and where a new visible scalar $h_2$ is unconstrained by the direct collider searches. On the left panel, we show the $Z^\prime$ production cross section times its branching ratio to the first- and second-generation leptons, $\sigma B \equiv \sigma\(pp \to Z^\prime\) \times B\(Z^\prime \to \ell \ell \) $ with $\ell = e,\mu$, as a function of the new vector boson mass and the new physics contribution to the muon anomalous magnetic moment $\Delta a^\ro{NP}_\mu$ (colour scale). On the right panel, we show the new scalar mass as a function of the same observables. All points above 
the solid line are excluded at $95\%$ C.L.~by the limit on $Z^\prime$ direct searches at the LHC performed by the ATLAS experiment \cite{Aad:2019fac} and are represented in grey shades. Darker shades denote would-be-scenarios with larger values of $\Delta a^\ro{NP}_\mu$ while the smaller contributions to this observable are represented with the lighter shades. The red cross in our figures signals the lightest $Z^\prime$ found in our scan which we regard as a possible early-discovery (or early-exclusion) benchmark point in the forthcoming LHC runs. Such a benchmark point is shown in the first line of Tab.~\ref{tab:bench}. On the right panel, we notice that the new scalar bosons can become as light as $400~\ro{GeV}$, with $Z^\prime$ masses being above $5.2~\ro{TeV}$. Such a moderately large minimal value for the new Higgs boson mass results from the fact that both the $h_2$ and the $Z'$ bosons share a common VEV in their mass forms as seen in Eqs.~\eqref{eq:simplify} and \eqref{eq:mZ}. Then, while direct searches at the LHC for a B-L-SM $Z'$ boson keep pushing its mass to larger values, the new Higgs boson mass also increases linearly with $m_{Z'}$ according to
\begin{equation}
	m_{h_2} \approx \sqrt{\frac{\lambda_2}{2}} \frac{m_{Z'}}{g_\mathrm{B-L}}\,. \label{mh2-lambda2}
\end{equation} 
Furthermore, neither $\lambda_2$ can be arbitrarily small (see Fig.~\ref{fig:STU-lambda-excl} central panels), not $g_\mathrm{B-L}$ can be arbitrarily large (see Fig.~\ref{fig:STU-gBL_Zp_HP} left panels) in order to compensate an increase in $m_{Z'}$. We highlight with a cyan diamond the benchmark point with the lightest $h_2$ boson within this range. This point is shown in the second line of Tab.~\ref{tab:bench}.
\begin{figure}[!htb]
	\centering
	\includegraphics[scale=0.37]{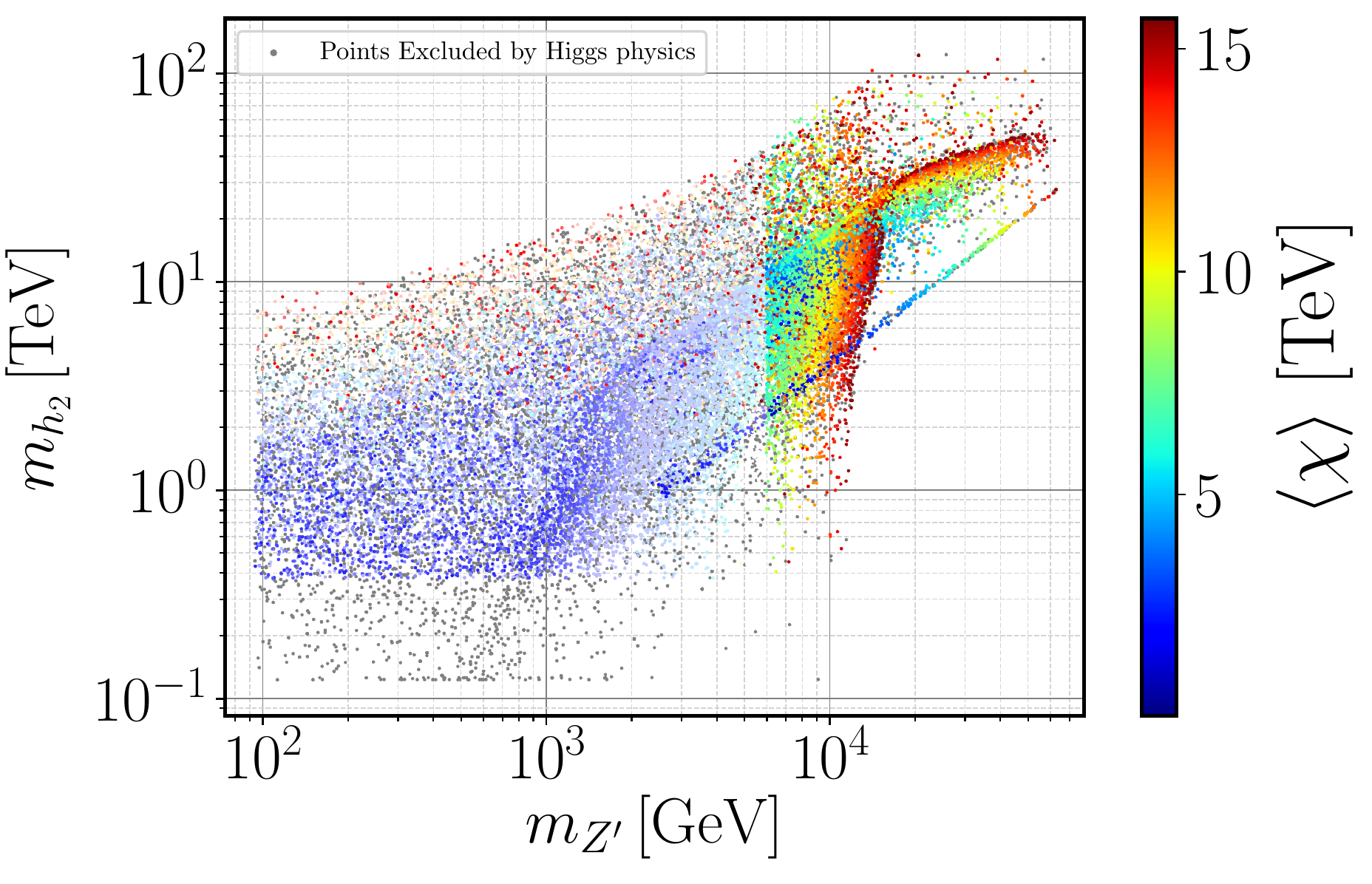}
	\includegraphics[scale=0.37]{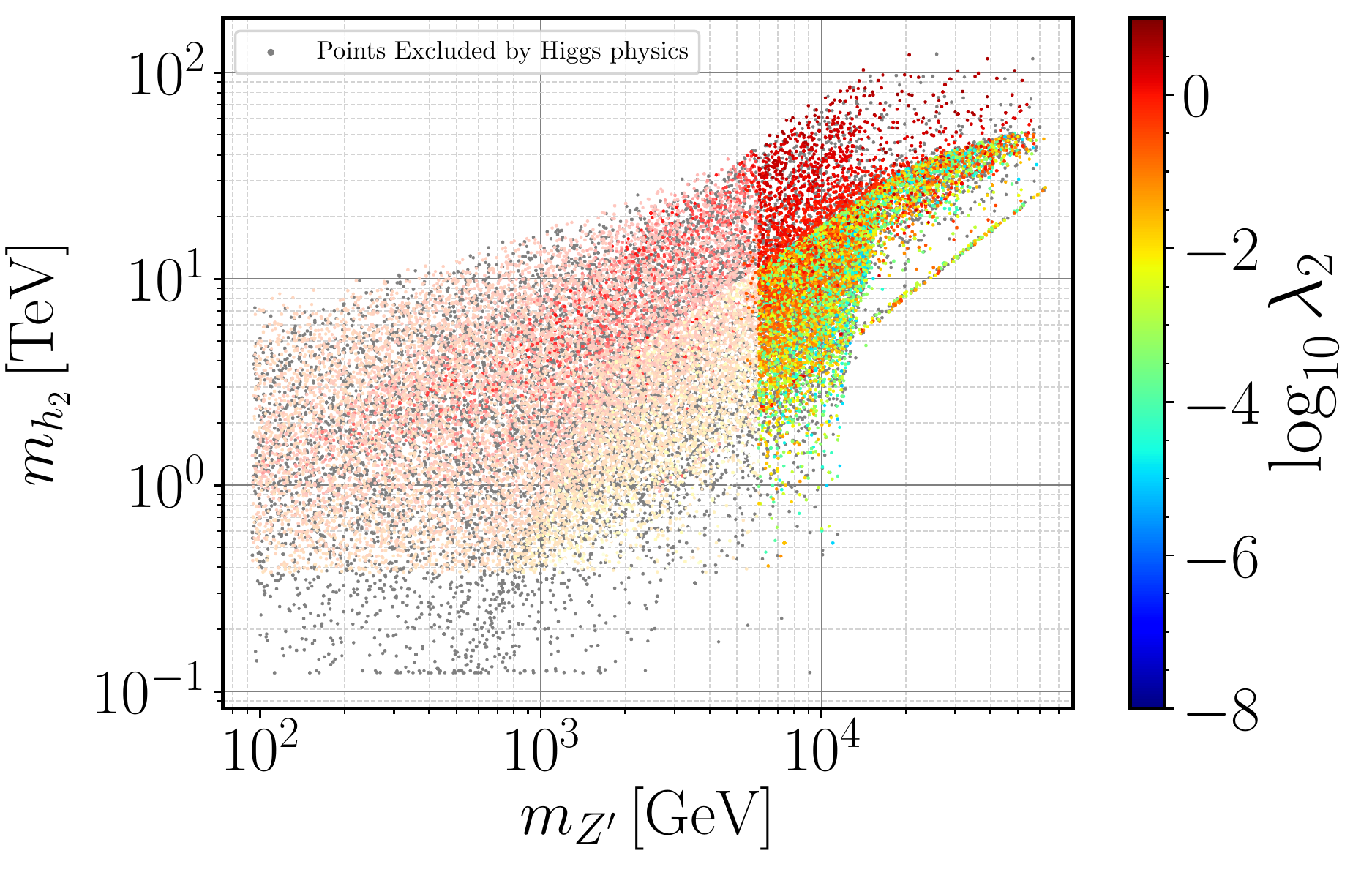}
	\caption{Scatter plots showing the scalar VEV (left) and the $\lambda_2$ coupling (right) versus the second Higgs mass and the $Z'$ mass. All the shown points 
	pass the theoretical and phenomenological constraints except the Higgs physics constraints (grey points) and ATLAS $Z'$ search constraints (faded colored points),
	while the colored points are the physically valid points that pass all the exclusion limits.}
	\label{fig:Plots-Higgs-physics}
\end{figure}	

The same observation can also be made from Fig.~\ref{fig:Plots-Higgs-physics} which represent the points excluded by Higgs physics constraints and by ATLAS $Z'$ search constraints as well as passed 
physically valid points in our numerical scan. The left panel shows the the scalar VEV versus the second Higgs mass and the $Z'$ mass, while the right panel illustrates
such dependence for $\lambda_2$ coupling roughly related to $m_{h_2}$ by means of Eq.~(\ref{mh2-lambda2}). Indeed, we observe that due to the Higgs physics constraints, $\lambda_2$ 
cannot be arbitrarily small in order to compensate a larger $Z’$ mass (thus, a large singlet VEV). In particular, the smaller the VEV or $m_{Z’}$, the larger $\lambda_2$. Therefore, the Higgs searches puts strongest constraints on $m_{h_2}$, at least, in the case of large $Z’$ mass.
\begin{table}[htb!]
\begin{center}
		\begin{tabular}{cccccccccc}
			$m_{Z^\prime}$ & $m_{h_2}$ &  $x$& $ \log_{10} \Delta a_\mu^\ro{NP}$ & $\sigma B$ & $\theta_W^\prime$ & $\log_{10}\alpha_h$ & $\g{B-L}{}$ & $\g{YB}{}$ & 
			$ \g{L}{\ell \ell Z^\prime} = \g{R}{\ell \ell Z^\prime}$ 
			\vspace{1mm}
			\\
			\hline \vspace{-1mm} \\ 
$5.199$    & $6.41$     & $15.4$    & $-13.01$       & $1.16 \times 10^{-5}$   & $\approx 0$                    & $-5.18$     & $0.17$   & $2.0\times 10^{-5}$    & $0.08$  
\vspace{1mm}  \\ 
$6.478$    & $0.41$     & $9.77$    & $-12.57$       & $2.15 \times 10^{-5}$   & $3.22\times 10^{-7}$      & $-5.85$     & $0.34$   & $1.7\times 10^{-3}$    & $0.17$  
\vspace{1mm}   \\ 
$6.371$    & $2.34$     & $1.08$   & $-11.05$        & $0.01$                           & $1.05\times 10^{-6}$      & $-7.31$     & $1.97$   & $2.1\times 10^{-3}$    & $0.98$   
\vspace{1mm}    \\ 
$6.260$    & $2.31$     & $1.15$   & $-11.07$        & $0.01$                           & $5.87\times 10^{-5}$      & $-2.79$     & $1.87$   & $0.125$                        & $0.94$  
\vspace{1mm}   \\ 
$6.477$    & $2.40$      & $1.14$   & $-11.08$       & $0.01$                           & $2.75\times 10^{-5}$      & $-4.29$     & $1.93$   & $0.06$                          & $0.97$  
\vspace{1mm}    \\ 
$6.252$    & $2.53$     & $1.28$   & $-11.08$        & $0.01$                           & $\approx 0$                    & $-8.65$     & $1.86$   & $1.6\times 10^{-5}$     & $0.93$   
\vspace{1mm}   \\ 
\end{tabular}
		\caption{A selection of five benchmark points represented in Figs.~\ref{fig:Plots1} and \ref{fig:Plots4} to \ref{fig:Plots2}. The $m_{Z^\prime}$, $m_{h_2}$ and $x$ parameters are given in TeV. The second line represents a point with lightest possible $h_2$ while the first one shows the lightest allowed $Z^\prime$ boson found in our scan. The last four lines show four points that yield 
a minimal tension $3.28$ standard deviations, with the combined theoretical and experimental error of the muon $(g-2)_\mu$ anomaly.}
		\label{tab:bench}
\end{center}
\end{table}

\subsubsection{\it Implications of direct $Z^\prime$ searches at the LHC for the $\left(g-2\right)_\mu$ anomaly}

Looking again to Fig.~\ref{fig:Plots1} (left panel), we see that there is a dark-red region where $\Delta a_\mu^\ro{NP}$ can be enhanced up to a maximum of $\Delta a_\mu^\mathrm{NP} = 8.9 \times 10^{-12}$ for a range of $m_{Z^\prime}$ boson masses approximately between $6.3~\ro{TeV}$ and $6.5~\ro{TeV}$, representing a very small improvement in comparison to the SM prediction. Such a mass region is particularly interesting as it can be probed by the forthcoming LHC runs. If a $Z^\prime$ boson discovery remains elusive, it can exclude a possibility of alleviating the tension between the measured and the SM prediction for the muon $\left(g-2\right)_\mu$ anomaly in the context of the B-L-SM. However, note that with new measurements at the E989 experiment at Fermilab, if a partial reduction in the current discrepancy is observed, the B-L-SM prediction may become an important result and a motivation for future $Z'$ searches at the LHC. Note, such maximal $\Delta a^\ro{NP}_\mu$ values represent a rather small region of scattered
red points where the new scalar boson mass takes values of a few TeV. Furthermore, in some scenarios represented by the second and third lines in Tab.~\ref{tab:bench}, the $\theta_W^\prime$ and $\alpha_h$ angles are not vanishingly small which may hint certain possibilities for observing both a new scalar and new vector boson in this region.
\begin{figure}[!htb]
	\centering
	\includegraphics[scale=0.75]{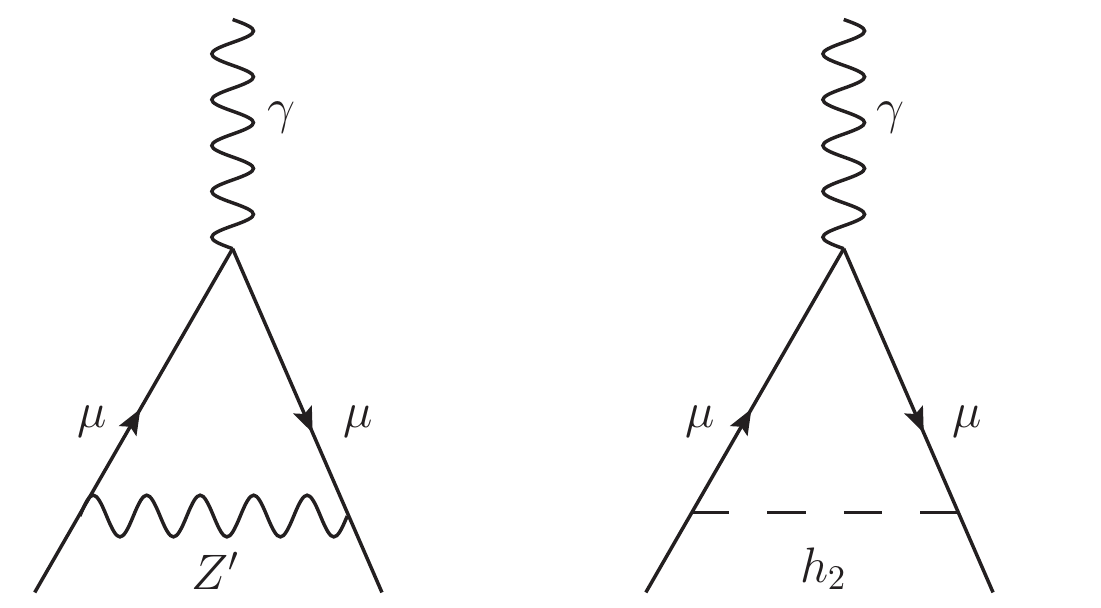}
	\caption{One-loop diagrams contributing to $\Delta a_\mu^\ro{NP}$ in the B-L-SM.}
	\label{fig:g-2}
\end{figure}	

\begin{figure}[!htb]
	\centering
	\includegraphics[scale=0.37]{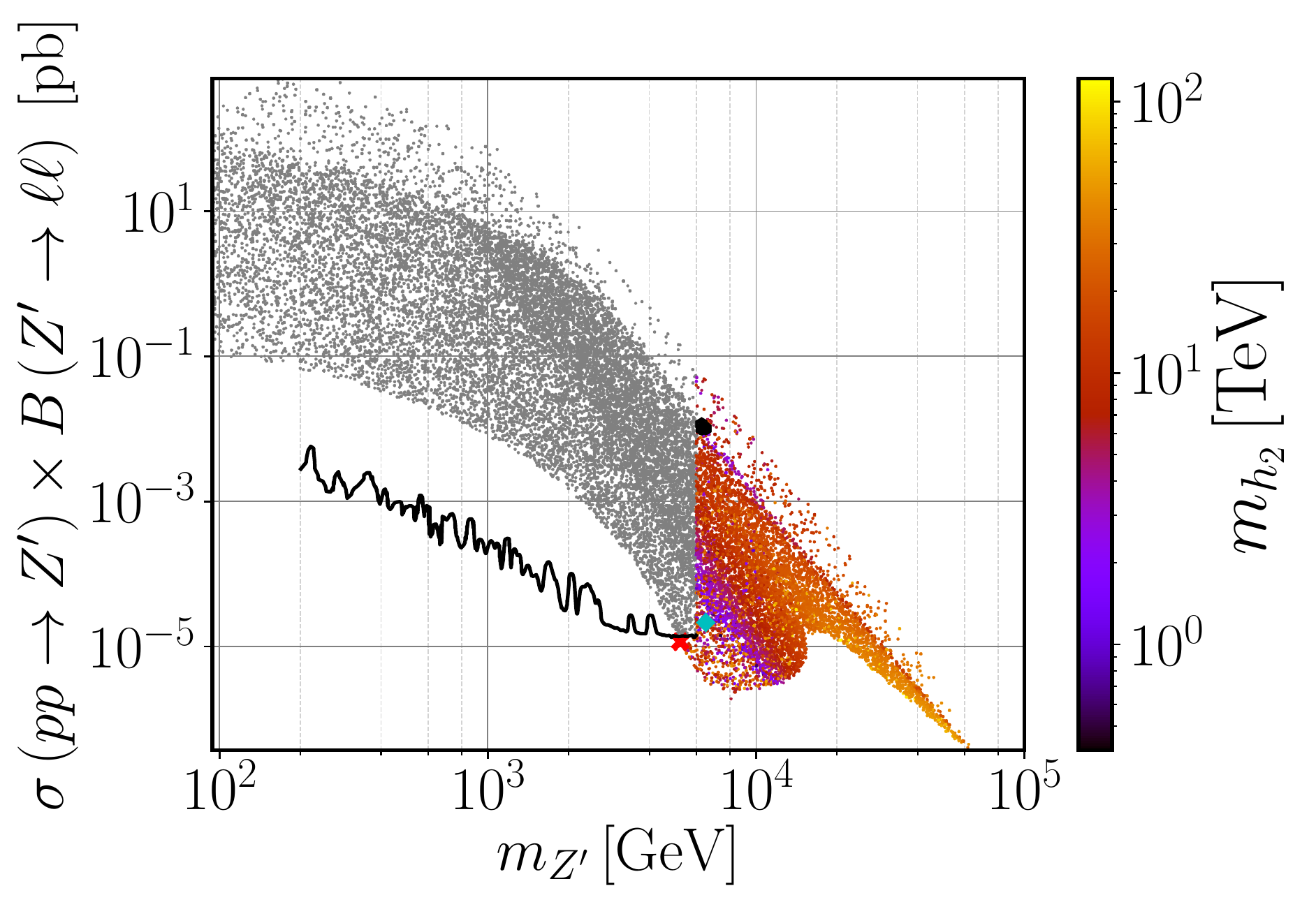}
	\includegraphics[scale=0.37]{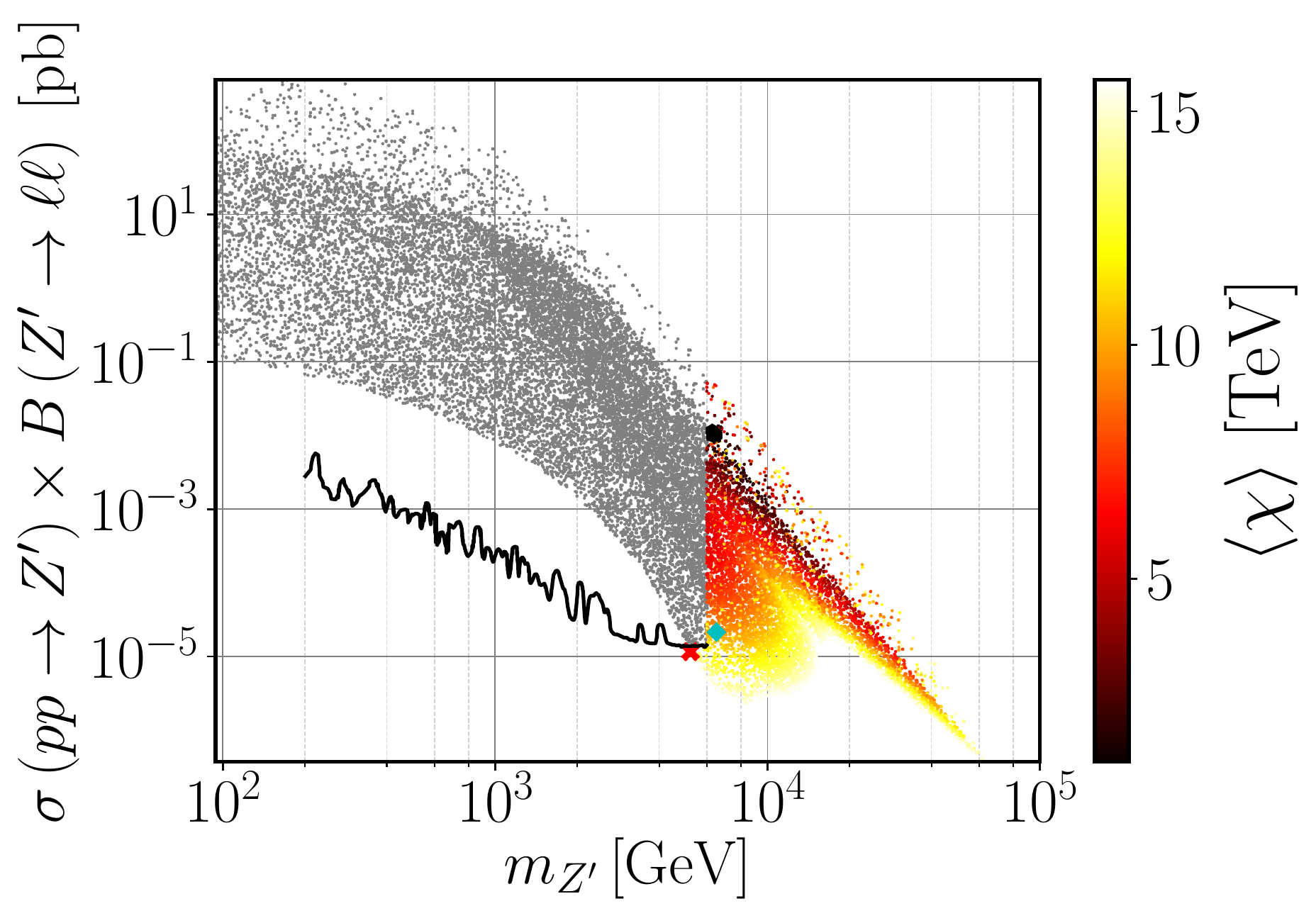}
	\includegraphics[scale=0.37]{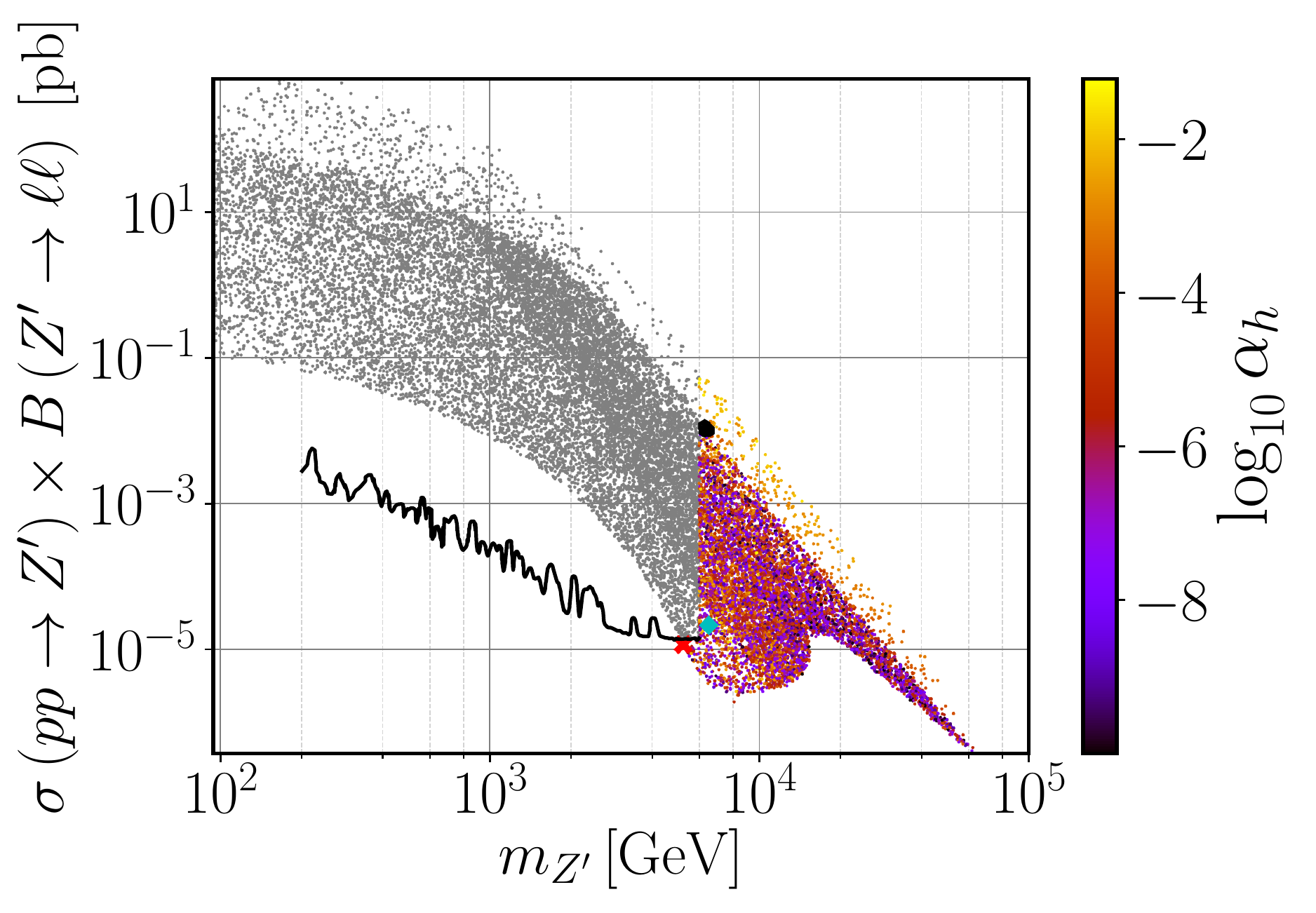}	
	\caption{Scatter plots showing the $Z^\prime$ Drell-Yan production cross section times the decay branching ratio into a pair of electrons and muons in terms of the $m_{Z^\prime}$ boson mass. The colour gradation represents the new scalar mass (top-left), the $\U{B-L}$-breaking VEV (top-right) and the scalar mixing angle (bottom). The notation is the same as in Fig.~\ref{fig:Plots1} (left). }
	\label{fig:Plots4}
\end{figure}

New physics contributions $\Delta a_\mu^\ro{NP}$ to the muon anomalous magnetic moment are given at one-loop order by the Feynman diagrams depicted in Fig.~\ref{fig:g-2}.
Since the couplings of a new scalar $h_2$ to the SM fermions are suppressed by a factor of $\sin \alpha_h$, which we find to be always smaller than $0.002$ as can be seen in the bottom panel of Fig.~\ref{fig:Plots4}, the right diagram in Fig.~\ref{fig:g-2}, which scales as $\Delta a_\mu^{h_2} \propto \tfrac{m_\mu^2}{m_{h_2}^2}\(y_\mu \sin \alpha_h\)^2$ with $y_\mu = Y_e^{22}$, provides sub-leading contributions to $\Delta a_\mu$. Furthermore, as we show in the top-left panel of Fig.~\ref{fig:Plots4} the new scalar boson mass, which we have found to satisfy $m_{h_2} \gtrsim 400~\ro{GeV}$, is not light enough to compensate the smallness of the scalar mixing angle. Conversely, and recalling that all fermions in the B-L-SM transform non-trivially under $\U{B-L}$, the new $Z^\prime$ boson can have sizeable couplings to fermions via gauge interactions proportional to $\g{B-L}{}$ and $\g{YB}{}$, essentially constrained by four fermion contact interactions. 

\begin{figure}[!htb]
	\centering
	\includegraphics[scale=0.37]{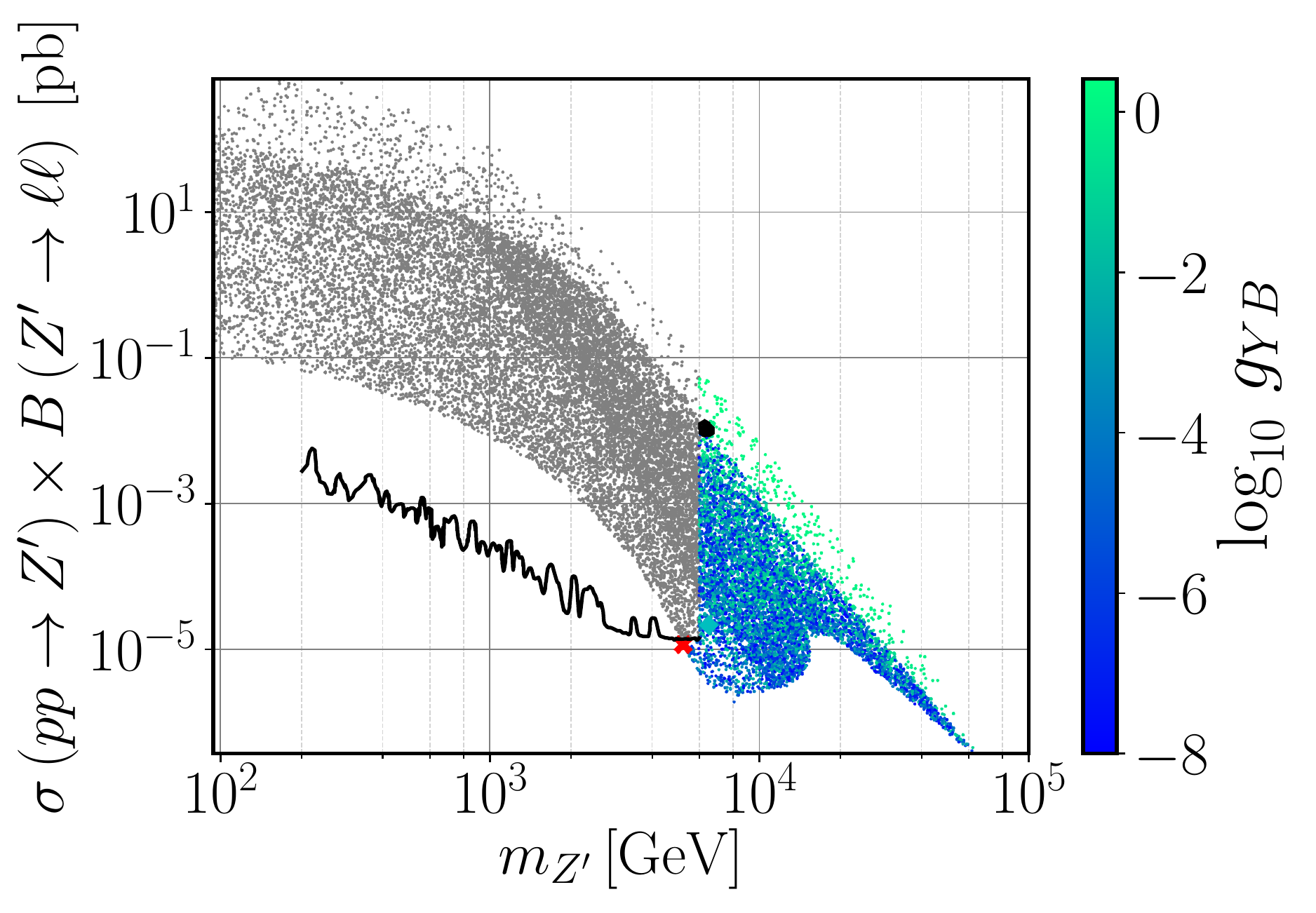}
	\includegraphics[scale=0.37]{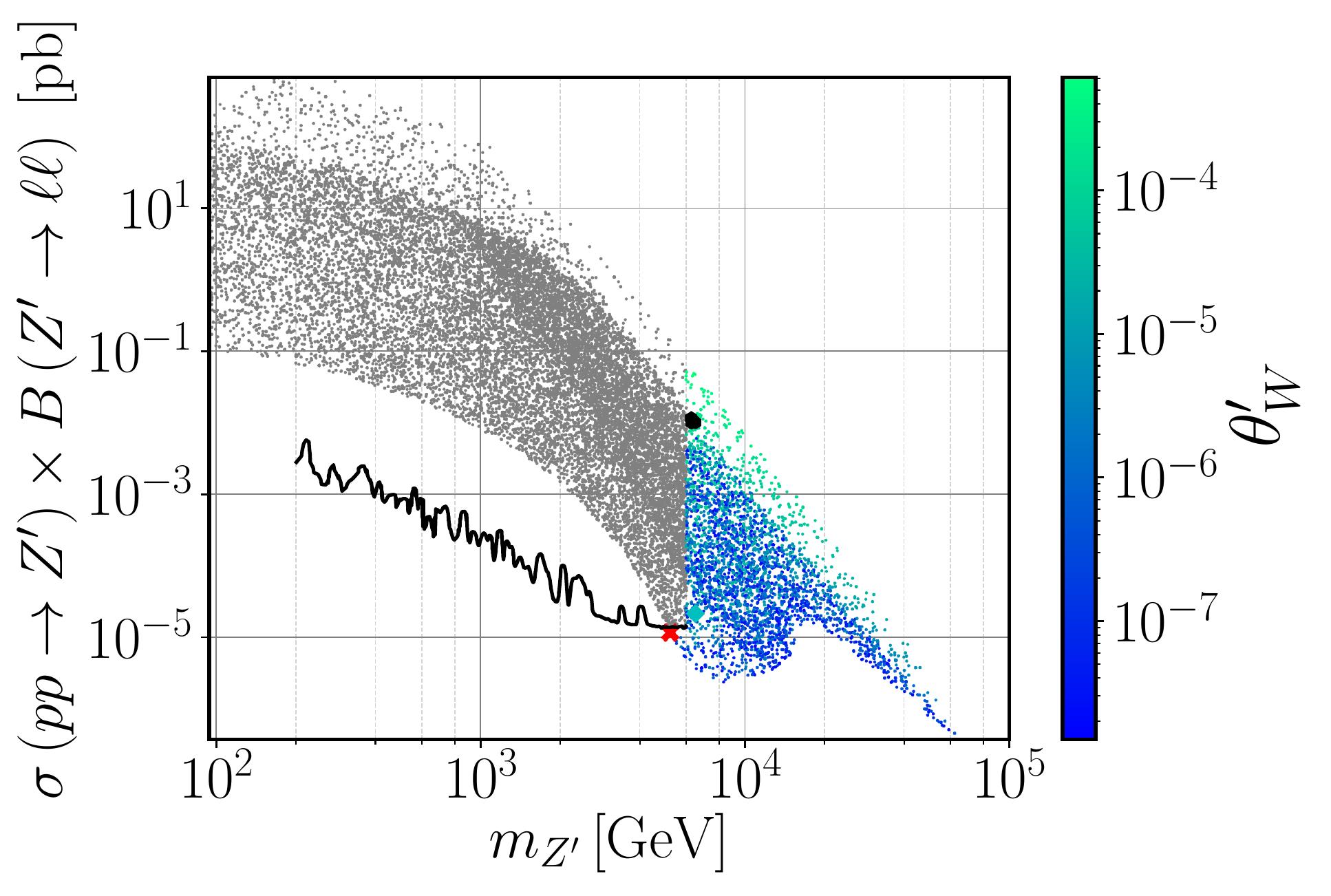}
	\includegraphics[scale=0.37]{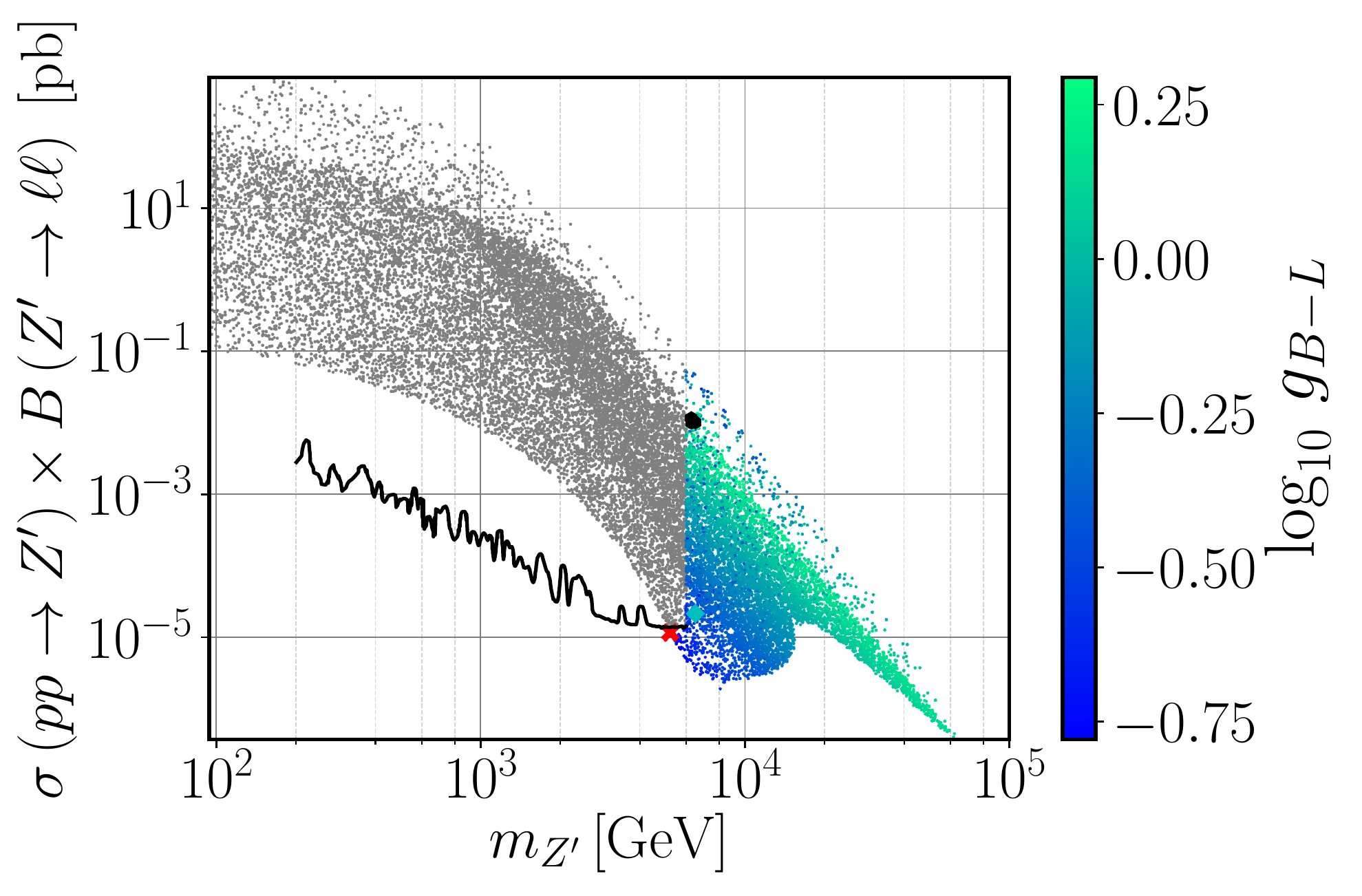}	
	\caption{The same as in Fig.~\ref{fig:Plots4} but with the colour scale representing the gauge-mixing parameters $\g{YB}{}$ (top-left) 
	and $\theta_W^\prime$ (top-right), and the $\U{B-L}$ gauge coupling, $\g{B-L}{}$ (bottom).}
	\label{fig:Plots3}
\end{figure}

\begin{figure}[!htb]
	\centering
	\includegraphics[scale=0.37]{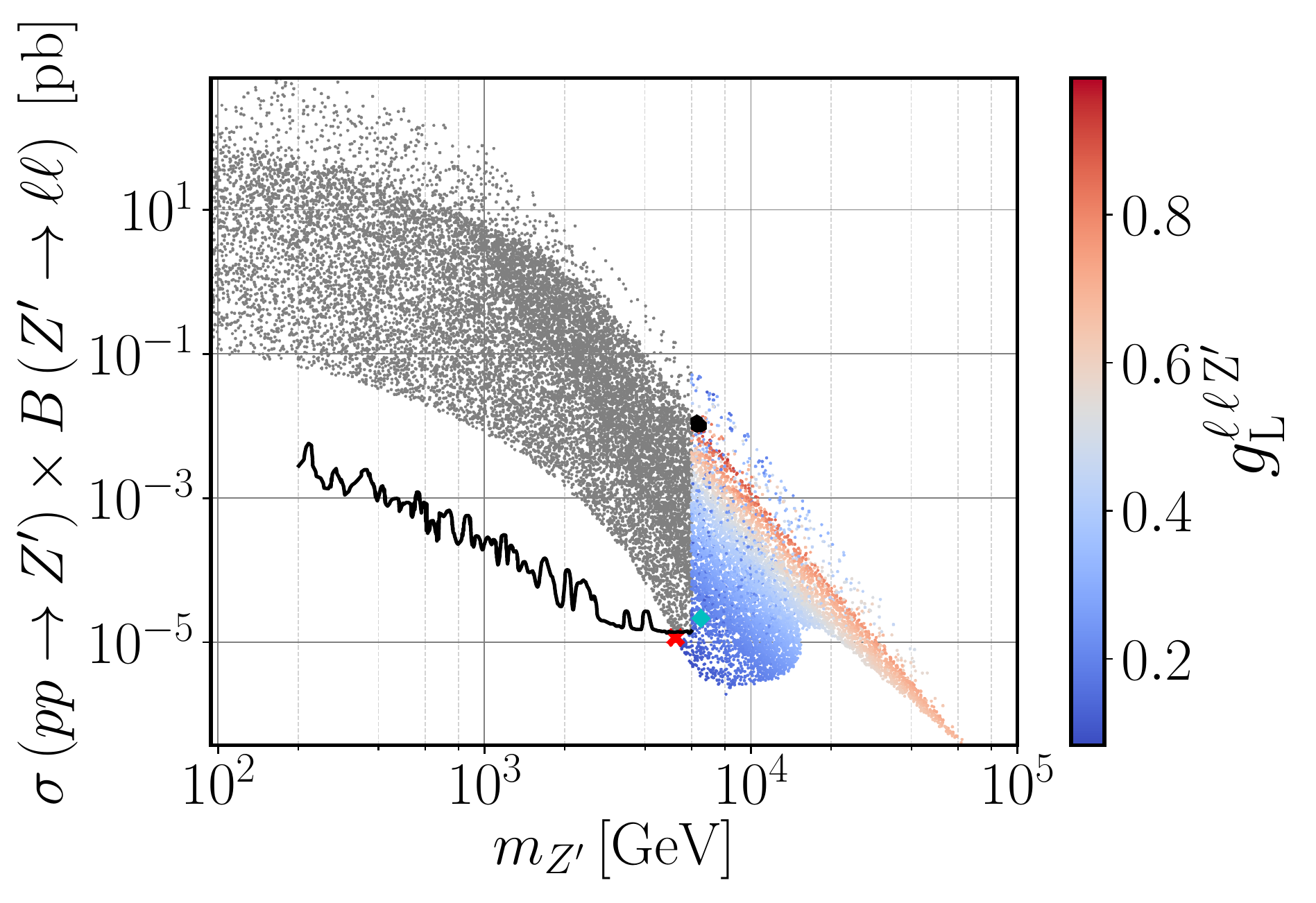}
	\includegraphics[scale=0.37]{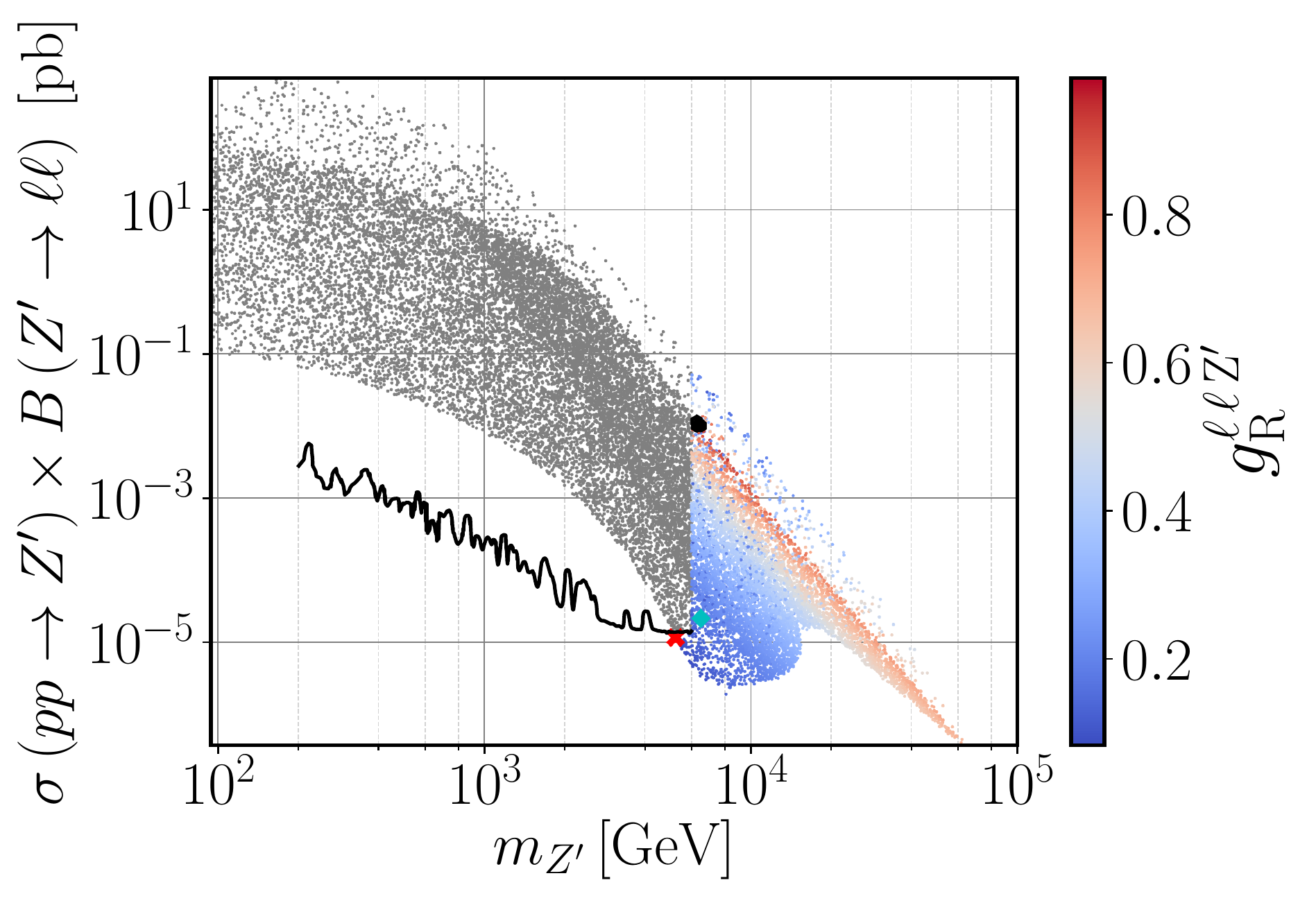}
	\includegraphics[scale=0.37]{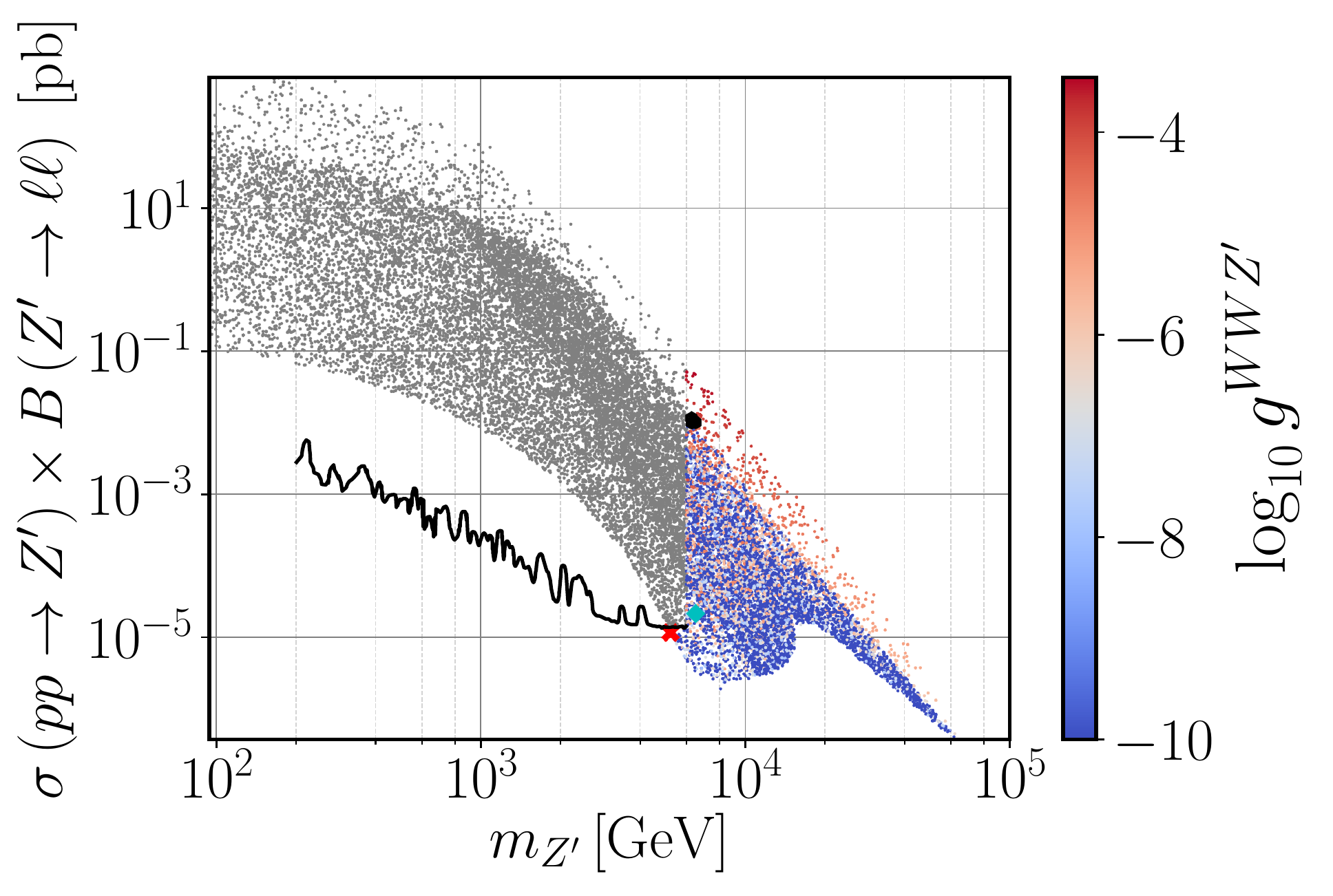}
    \includegraphics[scale=0.37]{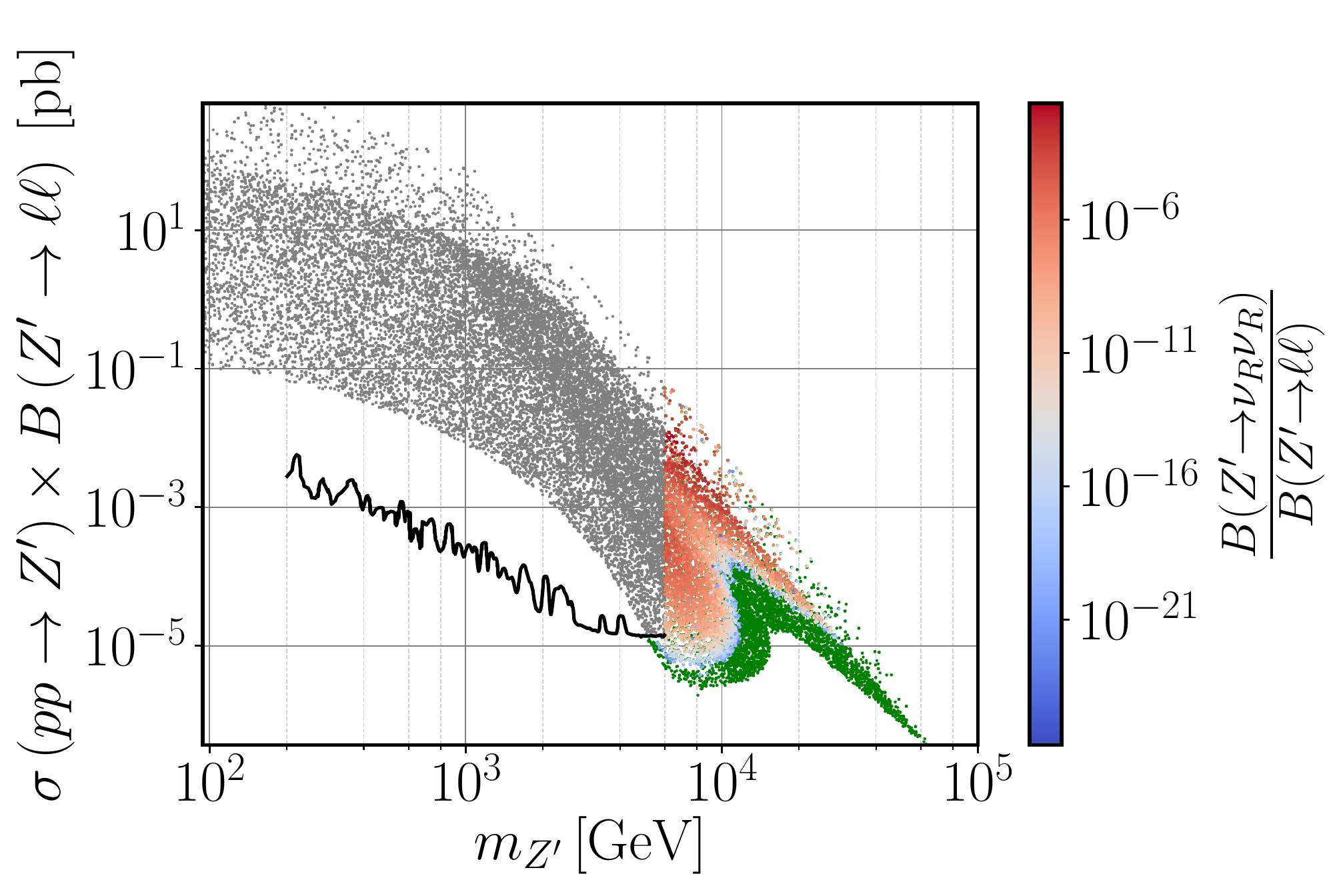}
	\caption{The same as in Fig.~\ref{fig:Plots4} but with the colour scale representing the coupling of leptons to the $Z^\prime$ (top panels), 
	the coupling of $W$ bosons to $Z^\prime$ (bottom left) and the ratio of $Z^\prime$ branching 
	fraction to right-handed neutrinos $\nu_R$ over its branching fraction to charged leptons (bottom right). 
	In the bottom-right panel, green points correspond to allowed scenarios with fixed $B(Z^\prime \to \nu_R \nu_R)=0$. }
	\label{fig:Plots2}
\end{figure}

Therefore, the left diagram in Fig.~\ref{fig:g-2} provides the leading contribution to the $\left(g-2\right)_\mu$ in the model under consideration.
In particular, $\Delta a_\mu^{Z^\prime}$ can be written as
\begin{equation}
\Delta a_\mu^{Z^\prime} = \tfrac{1}{4 \pi^2} \tfrac{m_\mu^2}{m_{Z^\prime}^2} \[\g{L}{\mu \mu Z^\prime} \g{R}{\mu \mu Z^\prime} g_\ro{FFV}\(\frac{m_\mu^2}{m_{Z'}^2}\) + \({\g{L}{\mu \mu Z^\prime}}^2 + {\g{R}{\mu \mu Z^\prime}}^2\) f_\ro{FFV}\(\frac{m_\mu^2}{m_{Z'}^2}\) \]
\label{eq:ZpContribution}
\end{equation}
where the left- and right-chiral projections of the charged lepton couplings to the $Z^\prime$ boson, $\g{L}{\ell \ell Z^\prime}$ and $\g{R}{\ell \ell Z^\prime}$, respectively, can be approximated as follows
\begin{equation}
\begin{aligned}
    \g{L}{\ell \ell Z^\prime} &\simeq \frac12\g{B-L}{} 
    +\frac14 \g{YB}{}
    + \frac{1}{32} \(\frac{v}{x}\)^2 \frac{\g{YB}{}}{\g{B-L}{2}} \(\g{Y}{2} - g^2 \)\,,
    \\
    \g{R}{\ell \ell Z^\prime} &\simeq \frac12\g{B-L}{}
    + \frac12\g{YB}{} 
    + \frac{1}{16} \(\frac{v}{x}\)^2 \frac{\g{YB}{}}{\g{B-L}{2}}\g{Y}{2}\,,
\end{aligned}\label{eq:gllZ}
\end{equation}
to the second order in $v/x$-expansion. The regions of the parameter space that we are exploring feature a heavy $Z'$ boson such that $m_\mu^2 \ll m_{Z'}^2$. 
In such a limit the loop functions $g_\ro{FFV}\(\tfrac{m_\mu^2}{m_{Z'}^2}\)$ and $f_\ro{FFV}\(\tfrac{m_\mu^2}{m_{Z'}^2}\)$tend to the values $g_\ro{FFV}\(0\) \to 4$ and $f_\ro{FFV}\(0\) \to -\tfrac{4}{3}$ where Eq.~\eqref{eq:ZpContribution} can be approximated to
\begin{equation}
\Delta a_\mu^{Z^\prime} \approx \tfrac{1}{3 \pi^2} \tfrac{m_\mu^2}{m_{Z^\prime}^2} \[3 \g{L}{\mu \mu Z^\prime} \g{R}{\mu \mu Z^\prime} - \({\g{L}{\mu \mu Z^\prime}}^2 + {\g{R}{\mu \mu Z^\prime}}^2\) \]\,.
\label{eq:ZpContribution-2}
\end{equation}
If $v/x \ll 1$, corresponding to the lighter shades of the color scale in the top-right panel of Fig.~\ref{fig:Plots4}, we can further approximate\footnote{ Even the yellow region in Fig.~\ref{fig:Plots4} has $v<x$ thus, expanding around small $v/x$ also provides a reliable approximation.}
\begin{equation}
\g{L}{\ell \ell Z^\prime} \simeq \frac12\(\g{B-L}{} + \dfrac{1}{2} \g{YB}{} \)\,,
\qquad
\g{R}{\ell \ell Z^\prime} \simeq \frac12\(\g{B-L}{} + \g{YB}{}\)\,. 
\label{eq:gLgR-simp}
\end{equation}
With this, the $Z'$ contribution to the muon anomalous magnetic moment can be recast as
\begin{equation}
\Delta a_\mu^{Z^\prime} \simeq \dfrac{1}{48 \pi^2} \dfrac{m_\mu^2}{m_{Z^\prime}^2} \[6 g_{_\ro{B-L}}^{} g_{_\ro{YB}}^{} + 4 g_{_\ro{B-L}}^{2} + g_{_\ro{YB}}^{2} \] \,,
\label{eq:amu-simple}
\end{equation}
and for $\g{YB}{} \ll \g{B-L}{}$, which represents the majority of the points in our scan,
\begin{equation}
\Delta a_\mu^{Z^\prime} \simeq \dfrac{1}{12 \pi^2} \dfrac{m_\mu^2}{m_{Z^\prime}^2} \g{B-L}{2}\,.
\label{eq:amu-simple2}
\end{equation}
Note that, limits from four fermion contact interactions do not allow $\g{B-L}{}$ to be sufficiently large to contribute to a sizeable $\Delta a_\mu^{\rm NP}$ via  Eqs.~\eqref{eq:amu-simple} or \eqref{eq:amu-simple2}. In particular, we found that $\g{B-L}{}$ is always smaller than $1.97$ as depicted in the bottom panel of Fig.~\ref{fig:Plots3}. On another hand, limits on the $\theta_W^\prime$ mixing angle and from four-fermion contact interactions do not forbid an order one $\g{YB}{}$ coupling in the sparser upper edge of the top-left panel of Fig.~\ref{fig:Plots3}. It is indeed such a sizeable $\g{YB}{}$ that only slightly enhances the muon $(g-2)_\mu$ anomaly as can be seen in the red region of both plots in Fig.~\ref{fig:Plots1}. We have found four benchmark points represented by the black dots in Figs.~\ref{fig:Plots1} and \ref{fig:Plots4} to \ref{fig:Plots2}, where the tension between the current combined $1 \sigma$ error of the muon anomalous magnetic moment and the B-L-SM prediction is alleviated only by at most $0.01$ standard deviations in comparison to the SM, a totally negligible effect. These points are shown in the third to the sixth lines of Tab.~\ref{tab:bench}.

A close inspection of Fig.~\ref{fig:Plots1} (left panel) and Fig.~\ref{fig:Plots4} (top-right panel) reveals an almost one-to-one correspondence between the colour shades. This suggests that $\Delta a_\mu^{Z^\prime}$ must somehow be related to the \vev~$x$. To understand this behaviour, let us also look to Fig.~\ref{fig:Plots3} (top-left panel) where we see that the coupling $\g{YB}{}$ is typically very small apart from the green band on the upper edge where it becomes of order one. For the relevant parameter space regions, Eq.~\eqref{eq:mZ} is indeed a good approximation as was argued above. It is then possible to eliminate $\g{B-L}{}$ from Eq.~\eqref{eq:amu-simple2} and rewrite it as
\begin{equation}
    \Delta a_\mu^{Z^\prime} \simeq \dfrac{y_\mu^2}{96 \pi^2} \(\dfrac{v}{x}\)^2 \qquad \text{for} \qquad \g{YB}{} \ll \g{B-L}{} \,,
    \label{eq:amu-vev}
\end{equation}
which explains the observed correlation between both Fig.~\ref{fig:Plots1} (left panel) and Fig.~\ref{fig:Plots4} (top-right panel).
Note that this simple and illuminating relation becomes valid as a consequence of the heavy $Z^\prime$ mass regime, in combination with the smallness of the $\theta_W^\prime$ mixing angle required by LEP constraints. Indeed, while we have not imposed any strong restriction on the input parameters of our scan (see Tab.~\ref{tab:scan}), Eq.~\eqref{eq:theta-p} necessarily implies that both $\g{YB}{}$ and $v/x$ cannot be simultaneously sizeable in agreement with what is seen in Fig.~\ref{fig:Plots3} (top-left panel) and Fig.~\ref{fig:Plots4} (top-right panel). The values of $\theta_W^\prime$ obtained in our scan are shown in the top-right panel of Fig.~\ref{fig:Plots3}.

For completeness, we show in Fig.~\ref{fig:Plots2} the physical couplings of $Z^\prime$ to muons (top panels) and to $W^\pm$ bosons (bottom-left panel).
Note that, for the considered scenarios, the latter can be written as
\begin{equation}
    g^{WWZ^\prime} \simeq \dfrac{1}{16} \dfrac{\g{YB}{}}{\g{B-L}{}} \(\dfrac{v}{x}\)^2\,.
    \label{eq:gWWZp}
\end{equation}
While both $\g{B-L}{}$ and the ratio $v/x$ provide a smooth continuous contribution in the $\sigma B - m_{Z^\prime}$ projection of the parameter space, the observed blurry region in $g^{WWZ^\prime}$ is correlated with the one in the top-left panel of Fig.~\ref{fig:Plots3} as expected from Eq.~\eqref{eq:gWWZp}. On the other hand, the couplings to leptons $g_{\rm L,R}^{\ell \ell Z^\prime}$ exhibit a strong correlation with 
$\g{B-L}{}$ in Fig.~\ref{fig:Plots3} except for the sparser region on the upper edge of the $\sigma B - m_{Z^\prime}$ plane where the correlation becomes proprotional to $\g{YB}{}$, in agreement with our discussion above and with Eqs.~\eqref{eq:gLgR-simp}. In the bottom-right panel of Fig.~\ref{fig:Plots2}, we have also shown the relative value of the $Z^\prime$ branching ratio into a pair of right-handed neutrinos, $B(Z^\prime \to \nu_R \nu_R)$, versus the corresponding branching fraction into charged leptons. We have found that the $Z^\prime$ decay into right-handed neutrinos is strongly suppressed for all the points that pass the theoretical and experimental constraints and thus cannot provide a significant impact on the exclusion bounds.

\subsubsection{\it Barr-Zee type contributions}
\label{sec:BarrZee}

\begin{figure}[!htb]
	\centering
	\includegraphics[scale=0.6]{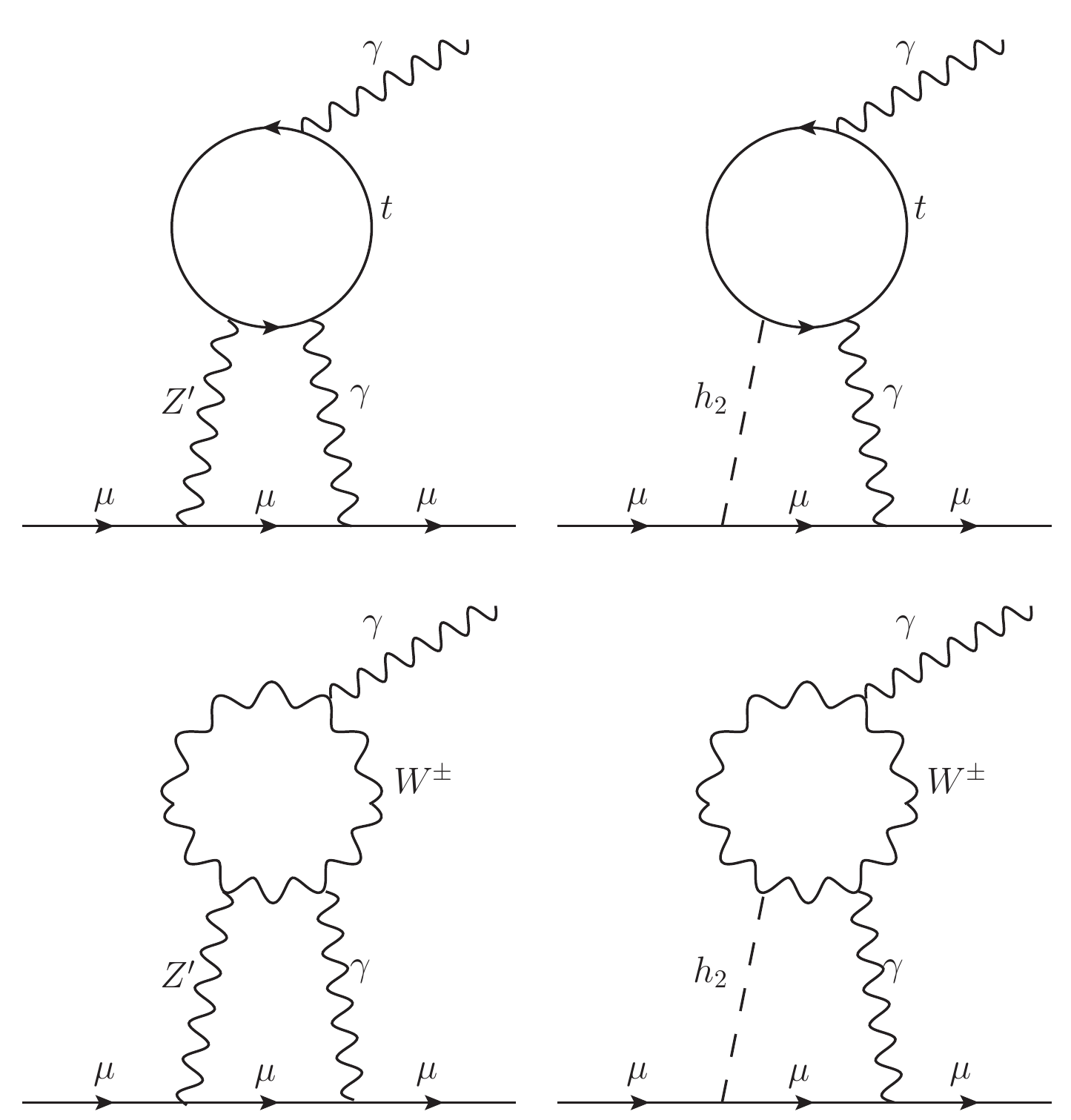}
	\caption{Barr-Zee type two-loop diagrams contributing to $\Delta a_\mu$.}
	\label{fig:Barr-Zee}
\end{figure}

To conclude our analysis, one should note that the two-loop Barr-Zee type diagrams \cite{Barr:1990vd} are always sub-dominant in our case. To see this, let us consider the four diagrams shown in Fig.~\ref{fig:Barr-Zee}.
The same reason that suppresses the one-loop $h_2$ contribution in Fig.~\ref{fig:g-2} is also responsible for the suppression of both the top-right and bottom-right diagrams in Fig.~\ref{fig:Barr-Zee} (for details see e.g.~Ref.~\cite{Ilisie:2015tra}). Recall that the coupling of $h_2$ to the SM particles is proportional to the scalar mixing angle $\alpha_h$, which is always small (or very small) as we can see in Fig.~\ref{fig:Plots4}. An analogous effect is present in the diagram involving a $W$-loop, where a vertex proportional to $g^{WWZ^\prime}$ suppresses such a contribution. The only diagram that might play a sizeable role is the top-left one where the couplings of $Z^\prime$ to both muons and top quarks are not negligible.

Let us then estimate the size of the first diagram in Fig.~\ref{fig:Barr-Zee}. This type of diagrams were already calculated in Ref.~\cite{Feng:2009gn} but for the case of a SM $Z$-boson. Since the same topology holds for the considered case of B-L-SM too, 
if we trade $Z$ by the new $Z^\prime$ boson, the contribution to the muon $(g-2)_\mu$ anomaly can be rewritten as
\begin{equation}
    \Delta a_\mu^{\gamma Z^\prime} = -\dfrac{g^2 \g{B-L}{2} m_\mu^2 \tan^2{\theta_W^\prime}}{1536 \pi^4} \(g_\ro{L}^{ttZ^\prime} - g_\ro{R}^{ttZ^\prime}\) \ro{T}_7\( m_{Z^\prime}^2, m_t^2, m_t^2 \)\,,
    \label{eq:agZ}
\end{equation}
where $g_\ro{L,R}^{ttZ^\prime}$, calculated in \texttt{SARAH}, are the left- and right-chirality projections of the $Z^\prime$ coupling to top-quarks, given by
\begin{equation}
\begin{aligned}
    g_\ro{L}^{ttZ^\prime} &= 
    - \dfrac{\g{YB}{}}{12} \cos{\theta_W^\prime}
    -\dfrac{\g{B-L}{}}{6} \cos{\theta_W^\prime} + \dfrac{g}{4} \cos{\theta_W} \sin{\theta_W^\prime} - \dfrac{\g{Y}{}}{12} \sin{\theta_W} \sin{\theta_W^\prime}\,,
    \\
    g_\ro{R}^{ttZ^\prime} &= 
    - \dfrac{\g{YB}{}}{3} \cos{\theta_W^\prime}
    -\dfrac{\g{B-L}{}}{6} \cos{\theta_W^\prime} - \dfrac{ \g{Y}{}}{3} \sin{\theta_W} \sin{\theta_W^\prime}\,.
\end{aligned}
\end{equation}
The loop integral $\ro{T}_7 \(m_{Z^\prime}^2, m_t^2, m_t^2\)$ was determined in Ref.~\cite{Feng:2009gn} and, in the limit $m_{Z^\prime} \gg m_t$, as we show in Eq.~\eqref{eq:T7-expanded}, it gets simplified to
\begin{equation}
    \ro{T}_7 \(m_{Z^\prime}^2, m_t^2, m_t^2\) \simeq \frac{2}{m_{Z^\prime}^2} \,,
    \label{eq:T7}
\end{equation}
up to a small truncation error (see Appendix~\ref{app:T7} for details). For the parameter space region under consideration the difference $g_\ro{L}^{ttZ^\prime} - g_\ro{R}^{ttZ^\prime}$ can be cast in a simplified form as follows 
\begin{equation}
    \(g_\ro{L}^{ttZ^\prime} - g_\ro{R}^{ttZ^\prime}\) \simeq \dfrac{1}{32} \g{YB}{} \[ 8 + \dfrac{\(g^2 + \g{Y}{2}\)}{\g{B-L}{2}} \(\dfrac{v}{x}\)^2 \] \approx \dfrac{1}{4} \g{YB}{} \,.
    \label{eq:gLminusgR}
\end{equation}
Using this result and the approximate value of the loop factor, we can calculate the ratio between 
the Barr-Zee type and one-loop contributions to the muon $(g-2)_{\mu}$,
\begin{equation}
    \dfrac{\Delta a_\mu^{\gamma Z^\prime}}{\Delta a_\mu^{Z^\prime}} \simeq -\dfrac{1}{4096 \pi^2}\dfrac{g^2 \(g^2 + \g{Y}{2}\) \g{YB}{3}}{\[6 \g{B-L}{} \g{YB}{}  + 4\g{B-L}{2} + \g{YB}{2} \] } \(\dfrac{v}{x}\)^4 \ll 1 \,,
\end{equation}
which shows that $\Delta a_\mu^{\gamma Z^\prime}$ does indeed play a subdominant role in our analysis and can be safely neglected.

\section{Conclusion}
\label{sec:Conclusions}

To summarise, in this work we have performed a detailed phenomenological analysis of the
minimal $\U{B-L}$ extension of the Standard Model known as the B-L-SM. In particular, 
we have confronted the model with the most recent experimental bounds from the direct 
$Z^\prime$ boson and next-to-lightest Higgs state searches at the LHC, as well as with 
the LEP constraints on four-fermion contact interactions. Simultaneously, 
we have analysed the prospects of the B-L-SM for a better explanation 
of the observed anomaly in the muon anomalous magnetic moment $(g-2)_{\mu}$ in comparison to the SM. 
For this purpose, we have explored the B-L-SM potential for interpretation of the $(g-2)_{\mu}$ 
anomaly in the regions of the model parameter space that 
are consistent with current constrains from the direct searches and electroweak precision observables.

We have studied the correlations of the $Z^\prime$ production cross section times the branching ratio 
into a pair of light leptons versus the physical parameters of the model. In particular, we have found 
that the muon $(g-2)_{\mu}$ observable dominated by $Z^\prime$ loop contributions is maximized 
for $m_{Z^\prime}$ between $6.3$ and $6.5~\ro{TeV}$. As one of the main results of our analysis, we have found phenomenologically 
consistent model parameter space regions that simultaneously fit the exclusion limits from direct $Z^\prime$ 
searches and maximize the muon $(g-2)_\mu$ contribution to a value of $8.9 \times 10^{-12}$. 
This represents a marginal or no improvement in comparison to the SM prediction. The new 
Muon $(g-2)_\mu$ E989 experiment at the Fermilab will be able to measure this anomaly with 
an increased precision of $0.14~\text{ppm}$. If a larger new physics contribution to this observable 
is confirmed, the B-L-SM can not be considered as a candidate theory to explain that effect.

One should notice here that the recent lattice result by the BMW collaboration \cite{Borsanyi:2020mff} 
suggests that there is no need for New Physics to explain the muon $(g-2)_\mu$ data.
If correct, this eliminates the necessarily large effects in the muon $(g-2)_\mu$ coming from the B-L-SM 
as compared to the SM. While a confirmation of the currently observed anomaly with a smaller error can 
become rather exciting news, a more pessimistic scenario when the discrepancy either disappears or partially 
reduces, would reinforce the significance of our result and offer a motivation for future $Z'$ searches at the LHC 
in the $5-7~\mathrm{TeV}$ domain. Along these lines, we have identified five benchmark 
points for future phenomenological explorations: one scenario with the lightest $Z^\prime$ 
($m_{Z^\prime}\simeq 5.2$ TeV), another scenario with the lightest second scalar boson ($m_{h_2}\simeq 400$ GeV), 
and three other scenarios that maximize the muon $(g-2)_{\mu}$ anomaly. 
Another important result resides in the fact that an increasingly heavy $Z'$ boson also pushes up 
the mass of the second Higgs boson. Therefore, the hypothetical observation of such new physics 
states as a scalar or a vector boson would pose stringent constraints on the B-L-SM.
For completeness, we have also estimated the dominant contribution from the Barr-Zee type 
two-loop corrections and found a relatively small effect.

To finalize, let us comment that with all most relevant constraints incorporated in our numerical analysis, while the best explanation of
the muon $(g-2)_\mu$ in the B-L-SM predicting a value marginally above the SM one is not satisfactory, our result offers an important 
piece of information that can be relevant for the upcoming $a_\mu$ precision measurements at Fermilab as well as for building less 
minimal models containing heavy $Z'$ bosons and capable of a good explanation of the muon $(g-2)_\mu$ anomaly.
Another research direction that can be taken is the BL-SM analysis for the conditions of the HL-LHC. Along these lines, 
a significance calculation in future $Z’$ searches similar to the one performed very recently in the 3-3-1 model in Ref.~\cite{Cogollo:2020afo}
that is probing the $Z’$ boson mass up to 4 TeV at the HL-LHC, should be pursued aiming at probing vast regions of 
the parameter space still allowed by the LHC searches.

\section*{Acknowledgments}

The authors would like to thank Werner Porod and Florian Staub for discussions on 
the \texttt{SPheno} implementation of the muon $(g-2)_{\mu}$. The authors would also like 
to thank Nuno Castro, Maria Ramos and Emanuel Gouveia for insightful discussions about the implementation 
of the current model in \texttt{MadGraph5\_aMC@NLO}. J.P.R~thanks Lund University for hospitality during a short and fruitful visit. J.P.R~is supported by the project PTDC/FIS-PAR/31000/2017. The work of A.P.M.~has been performed in the framework of COST Action CA16201 “Unraveling new physics at the LHC through the precision frontier” (PARTICLEFACE). A.P.M.~is supported by the Center for Research and Development in Mathematics and Applications (CIDMA) through the Portuguese Foundation for Science and Technology (FCT -Fundação para a Ciência e a Tecnologia), references UIDB/04106/2020 and UIDP/04106/2020 and by national funds (OE), through FCT, I.P., in the scope of the framework contract foreseen in the numbers 4, 5 and 6 of the article 23, of the Decree-Law 57/2016, of August 29, changed by Law 57/2017, of July 19. A.P.M. is also supported by the Enabling Green E-science for the Square Kilometer Array Research Infras-tructure(ENGAGESKA), POCI-01-0145-FEDER-022217, and by the projects PTDC/FIS-PAR/31000/2017, CERN/FIS-PAR/0027/2019 and CERN/FIS-PAR/0002/2019.~R.P.~is partially 
supported by the Swedish Research Council, contract number 621-2013-428.

\appendix

\section{The loop integral $\ro{T}_7(x,y,y)$}
\label{app:T7}

In Appendix B of Ref.~\cite{Feng:2009gn}, the exact integral equations for $\ro{T}_7\(x,y,z\)$ are provided. 
In our analysis we consider the limit where $x \gg y = z$, with $x = m_{Z^\prime}^2$ and $y = z = m_t^2$, 
where Eq.~\eqref{eq:T7} provides a good approximation up to a truncation error. Here, we show the main steps 
in determining Eq.~\eqref{eq:T7}. The exact form of the loop integral reads as
\begin{equation}
\begin{aligned}
    \ro{T}_7\(x,y,y\) =& -\dfrac{1}{x^2} \varphi_0\(y,y\) + 2 y \dfrac{\del^3 \Phi(x,y,y)}{\del x \del y^2} + \dfrac{\del^2 \Phi(x,y,y)}{\del x^2} + x \dfrac{\del^3 \Phi(x,y,y)}{\del x^2 \del y} \\
    & + \dfrac{\Phi(x,y,y)}{x^2}
    -\dfrac{1}{x} \dfrac{\del \Phi(x,y,y)}{\del x}
    + \dfrac{\del^2 \Phi(x,y,y)}{\del x \del y} \,,
\end{aligned}    
\label{eq:T7-Integrals}
\end{equation}
with $\varphi_0 (x,y)$ and $\Phi(x,y,z)$ defined in Ref.~\cite{Feng:2009gn}. Let us now expand 
each of the terms for $x \ll y$. While the first term is exact and has the form
\begin{equation}
    -\dfrac{1}{x^2} \varphi_0\(y,y\) = -2 \dfrac{y}{x^2} \log^2 y \,,
    \label{eq:expand1}
\end{equation}
the second can be approximated to
\begin{equation}
    2 y \dfrac{\del^3 \Phi(x,y,y)}{\del x \del y^2} \simeq \xi \dfrac{24}{x} = \dfrac{8}{x} ~\textrm{for}~ \xi = \dfrac{1}{3}\,.
    \label{eq:expand2}
\end{equation}
In Eq.~\eqref{eq:expand2}, the $\xi = \tfrac{1}{3}$ factor was introduced in order to compensate for a truncation error. This was obtained by comparing the numerical values of the exact expression and our approximation. The third term can be simplified to
\begin{equation}
    \dfrac{\del^2 \Phi(x,y,y)}{\del x^2} \simeq \dfrac{2}{x} \( \log y - \log \dfrac{y}{x} \) + \dfrac{2}{x} \,,
    \label{eq:expand3}
\end{equation}
and the fourth to
\begin{equation}
    x \dfrac{\del^3 \Phi(x,y,y)}{\del x^2 \del y} \simeq -\dfrac{4}{x}\(\log \dfrac{y}{x} + 1 \)\,.
    \label{eq:expand4}
\end{equation}
The fifth and the seventh terms read
\begin{equation}
    \dfrac{\Phi(x,y,y)}{x^2}
    -\dfrac{1}{x} \dfrac{\del \Phi(x,y,y)}{\del x} \simeq \dfrac{2}{x} \log \dfrac{1}{x} \,,
    \label{eq:expand5}
\end{equation}
and finally, the sixth terms can be expanded as
\begin{equation}
    \dfrac{\del^2 \Phi(x,y,y)}{\del x \del y} \simeq \dfrac{4}{x}\(\log \dfrac{y}{x} -1\)\,.
    \label{eq:expand6}
\end{equation}
Noting that Eq.~\eqref{eq:expand1} is of the order $\tfrac{1}{x^2}$, putting together Eqs.~\eqref{eq:T7-Integrals}, \eqref{eq:expand2}, \eqref{eq:expand3}, \eqref{eq:expand4}, \eqref{eq:expand5}, and \eqref{eq:expand6} we get for the leading $\tfrac{1}{x}$ contributions the following:
\begin{equation}
    \begin{aligned}
    \ro{T}_7\(x,y,y\) \simeq& \overbrace{\dfrac{2}{x} \( \log y - \log \dfrac{y}{x} \) + \dfrac{2}{x} \log \dfrac{1}{x}}^{0} \overbrace{-\dfrac{4}{x}\(\log \dfrac{y}{x} + 1 \) +
    \dfrac{4}{x}\(\log \dfrac{y}{x} -1\)}^{-\tfrac{8}{x}} \\
&+ \dfrac{8}{x} + \dfrac{2}{x}
\simeq \dfrac{2}{x}\,.
    \end{aligned}
    \label{eq:T7-expanded}
\end{equation}

\bibliographystyle{JHEP}

\bibliography{bib}


\clearpage

\end{document}